\documentclass[aps,prd,nofootinbib,twocolumn]{revtex4}
\usepackage{amssymb}
\usepackage{amsmath,color}
\usepackage{epsfig}
\usepackage{graphicx,epsf,epsfig}
\usepackage{bbm}
\usepackage{subfigure}
\usepackage{multirow}
\usepackage{setspace}
\usepackage{verbatim}
\usepackage{epstopdf}
\usepackage[linktocpage]{hyperref}
\usepackage{multirow}
\newcommand{\be}{\begin{eqnarray}}
\newcommand{\ee}{\end{eqnarray}}

\begin{document}

\title{Gravitational wave production after inflation with cuspy potentials}

\author{Jing Liu$^{1,2}$}
\email{liujing@itp.ac.cn}

\author{Zong-Kuan Guo$^{1,2}$}
\email{guozk@itp.ac.cn}

\author{Rong-Gen Cai$^{1,2}$}
\email{cairg@itp.ac.cn}

\author{Gary Shiu$^3$}
\email{shiu@physics.wisc.edu}

\affiliation{$^1$CAS Key Laboratory of Theoretical Physics, Institute of Theoretical Physics,
 Chinese Academy of Sciences, P.O. Box 2735, Beijing 100190, China}
\affiliation{$^2$School of Physical Sciences, University of Chinese Academy of Sciences,
 No.19A Yuquan Road, Beijing 100049, China}
\affiliation{$^3$Department of Physics, University of Wisconsin-Madison, Madison, WI 53706, USA}

\begin{abstract}
We investigate the effect of the cuspiness of scalar potentials on the production of gravitational waves
during oscillon formation after inflation.
We consider a more general form of potentials with a mass parameter $M$,
which repoduce cuspy potentials for fields much larger than $M$, and smooth potentials in the opposite limit.
For cuspy potentials, nonsmooth oscillations of the inflaton induce an amplification of the inflaton fluctuations at the bottom of the potential,
so that oscillons copiously form, which leads to a significant stochastic gravitational wave background with a double-peak spectrum.
By varying the parameter $M$, we
find that cuspy potentials yield stronger signals of gravitational waves 
and the generation of gravitational waves disappears for smooth potentials.
Moreover, we calculate the equation of state after inflation and find 
the presence of a quasi matter-dominated stage right before the transition to the radiation-dominated stage.
\end{abstract}

\maketitle

\section{Introduction}
Gravitational waves (GWs) play a distinctive role in the context of inflationary cosmology.
A stochastic background of GWs,
produced during inflation and subsequent preheating/reheating process after inflation,
carries useful information about the inflationary dynamics and
inflaton decay (see~\cite{Cai:2017cbj} for a recent review).
Detecting such a stochastic background of GWs, either directly or indirectly,
can provide us with a unique opportunity to test the inflationary scenario.

During inflation, quantum fluctuations of the scalar and tensor modes of the spacetime metric were stretched
by the accelerated expansion of the Universe, and were then nearly frozen on super-Hubble scales.
In the standard single-field slow-roll inflationary scenario,
the amplitude of the power spectrum of primordial tensor perturbations (i.e., GWs) produced during inflation
depends on the energy scale of inflation~\cite{Liddle:1993ch,Guo:2010mm}.
Since such GWs can result in the B-mode polarization of the cosmic microwave background (CMB) anisotropies,
their spectrum is, in principle, measurable by future CMB polarization experiments.
Although primordial GWs have not been detected yet,
the upper limits on the tensor spectrum (quantified by the tensor-to-scalar ratio $r$) have helped us to discriminate inflationary models.
Current CMB data alone already put an upper bound on the tensor-to-scalar ratio $r < 0.09$ at 95$\%$ confidence level~\cite{Array:2015xqh},
which have been effective in discriminating inflationary models in combination with the constraints on the scalar spectral index.
The Planck 2018 result is further tightened by combining with the BICEP2/Keck Array BK14 data to obtain $r_{0.002}<0.064$~\cite{Akrami:2018odb}.
For example, the models with cubic and quartic potentials are strongly disfavored,
and the quadratic potential is moderately disfavored by the Planck 2015 data~\cite{Ade:2015lrj},
while the axion monodromy inflation with a linear potential~\cite{McAllister:2008hb}
or fractional powers~\cite{Silverstein:2008sg} are compatible with the current Planck results.
Further advances in the axion monodromy inflation have suggested potentials with even more possible powers \cite{Marchesano:2014mla,McAllister:2014mpa}.
Moreover, it has recently been shown in~\cite{Landete:2017amp} that stringy effects
can lower the power of a quadratic axion monodromy potential to less than linear.
Thus, the axion monodromy inflation represents an interesting class of large-field inflationary models that are compatible with the CMB data.

In the inflationary scenario, another source of GWs is parametric resonance during preheating after inflation~\cite{Kofman:1994rk}.
During preheating,
the Fourier modes of a scalar matter field $\chi$ coupled to the inflaton
grow exponentially by parametric resonance, driven by the oscillating inflaton.
The modes are quickly pumped up to a large amplitude.
Such highly pumped modes correspond to large, time-dependent density inhomogeneities in configuration space,
ensuring that the matter distribution has a non-trivial quadrupole moment,
which can source significant GWs~\cite{Khlebnikov:1997di}.
It is found that the present peak frequency of such GWs is proportional to
the energy scale of inflation, while the present amplitude of GWs
is independent of the energy scale of inflation~\cite{Easther:2006gt,Easther:2006vd}.
If the inflaton is nonminimally coupled to the curvature, the coupling can enhance the peak value of the GW spectrum produced during preheating~\cite{Fu:2017ero,Zhu:2018smk}.
In hybrid inflation,
since the energy scale ranges from the GUT scales down to the TeV scale,
the stochastic background of GWs produced during preheating
is expected to be directly detected by future GW detectors~\cite{GarciaBellido:2007dg,GarciaBellido:2007af}.

Oscillons, localized non-topological quasi-solitons, can be generated during preheating~\cite{Broadhead:2005hn,Farhi:2007wj,Amin:2010dc,Gleiser:2011xj,Amin:2011hj,Lozanov:2016hid,Lozanov:2017hjm,Hasegawa:2017iay}
when the scalar potential satisfies the ``opening up'' condition~\cite{Amin:2010jq}.
In the oscillon preheating scenario, a stochastic background of GWs is produced when the oscillons are forming.
For a symmetric smooth potential, the GW production is not significant~\cite{Zhou:2013tsa},
while for an asymmetric smooth potential, oscillons can generate a peak in the energy spectrum of GWs,
which lies above the expected sensitivity cure of the fifth observing run of aLIGO-Virgo detector network~\cite{Antusch:2016con,Antusch:2017flz,Antusch:2017vga}.
Recently Ref.~\cite{Amin:2018xfe} showed that the dominant, growing high frequency peak in the asymmetric smooth potential
is a numerical artifact by using pseudo-spectral algorithms for numerical evaluation.
In models with a cuspy potential, the nonsmooth oscillations can trigger
amplification of fluctuations of the inflaton itself at the moment when $\phi(t)=0$,
so that oscillons copiously form during oscillations of the inflaton,
which sources a significant stochastic background of GWs~\cite{Liu:2017hua}.
Interestingly, these cuspy potentials lead to a characteristic energy spectrum of GWs with double peaks,
which can be distinguished from smooth potentials by measuring the shape of the energy spectrum of GWs.

In this paper we investigate the effect of the cuspiness of the potentials on the production of GWs from oscillons.
A class of potentials is adopted to mimic the cuspy potentials in the asymptotically smooth limit.
We find that the smoothness of the potentials near the point $\phi(t)=0$ suppresses the energy spectrum of GWs.
Moreover, we study the dynamics of oscillon formation and calculate the equation of state (EoS) parameter before radiation domination.

The paper is organized as follows. In Sec.~\ref{sec:model}, we briefly review the models we study in this paper.
In Sec.~\ref{sec:nummet}, we describe our numerical algorithm for the evolution of scalar fields and tensor perturbations in an expanding Universe.
In Sec.~\ref{sec:dyn}, we study the growth of linear perturbations with a semi-analytic method
and present our numerical studies of the nonlinear dynamics.
In Sec.~\ref{sec:stoGW}, we calculate the energy spectrum of GWs today.
Sec.~\ref{sec:condis} is devoted to conclusions and discussions.

\section{The Models}
\label{sec:model}
We consider a single-field inflationary model in which the inflaton is minimally coupled to gravity.
The action is given by
\begin{equation}
\label{eq:action}
S=\int d^{4}x \sqrt{-g}\left[-\frac{M_\mathrm{pl}^{2}}{2}R+\frac{1}{2}\partial_{\mu}\phi\partial^{\mu}\phi+V(\phi)\right]\,,
\end{equation}
where $M_\mathrm{pl}\equiv (8\pi G)^{-1/2}$ is the reduced Planck mass, $R$ is the Ricci scalar and $\phi$ is the inflaton.
The Planck team discussed inflationary models with cuspy potentials of the following form
in Ref.~\cite{Ade:2015lrj}
\begin{equation}
V(\phi)=\lambda M_\mathrm{pl}^{4-p}|\phi|^p,
\label{eq:potentials}
\end{equation}
with $p=1, \; 2/3,\; 2/5$.

In string theory, space-filling wrapped branes introduce an axion monodromy
that leads to a linear potential~\cite{McAllister:2008hb}.
Inflationary potentials proportional to $\phi^{2/3}$ and $\phi^{2/5}$ arise in compactifications on manifolds
with metric flux such as Nil manifolds which contain tori twisted over circles~\cite{Silverstein:2008sg}.
More generally, the monodromy generated by fluxes can lead to inflaton potentials with more varieties of powers~\cite{Marchesano:2014mla, McAllister:2014mpa}.
In the paper we hasten to add that the powers of these potentials are expected only at large field values, due to the
coupling of the inflaton to high-scale physics.
At the end of inflation, i.e., for small $\phi$, these potentials for the axion monodromy typically become quadratic. Nonetheless, cuspy potentials can arise in other inflationary contexts, e.g., through non-standard kinetic terms or as a result of integrating out the dynamics of other fields coupled to the inflaton.

Assuming the potential in Eq.~(\ref{eq:potentials}) applies to both the inflationary stage and at the end of inflation,
the parameter $\lambda$
in this simple class of models
can be fixed by the estimated amplitude of scalar perturbations
from the CMB data.
Assuming the number of $e$-folds $N=50$,
for powers of $p=1, 2/3, 2/5$, $\lambda \approx 3, 4, 5\times10^{-10}$,
the predicted scalar spectral index $n_s\approx 0.970, 0.973, 0.976$,
and the predicted tensor-to-scalar ratio $r\approx 0.08, 0.05, 0.03$,
respectively.
These predictions are consistent with the recent CMB data~\cite{Ade:2015lrj}.
In the preheating scenario, after inflation the inflaton $\phi$ begins to oscillate around the minimum of its potential
and ultimately decays into elementary particles in the standard model of particle physics.
However, for a cuspy potential, the oscillating behavior of the inflaton is very different from that of smooth potentials like $\phi^2$ and $\phi^4$.
It is found that an efficient parametric resonance can occur
during preheating for an inflaton potential~(\ref{eq:potentials}) with $0<p \leq 2$,
if the inflaton couples to a scalar matter field $\chi$ via an interaction term $\phi^2\chi^2$~\cite{Moghaddam:2015ava}.
Recently, the production of GWs has been studied during oscillations of the inflaton after inflation with the cuspy potentials~\cite{Liu:2017hua}.
The nonsmooth oscillations can trigger amplification of fluctuations of the inflaton itself,
so that oscillons copiously form, which leads to a characteristic energy spectrum of GWs with double peaks.

\begin{figure}[h]
\includegraphics[width=2.5in]{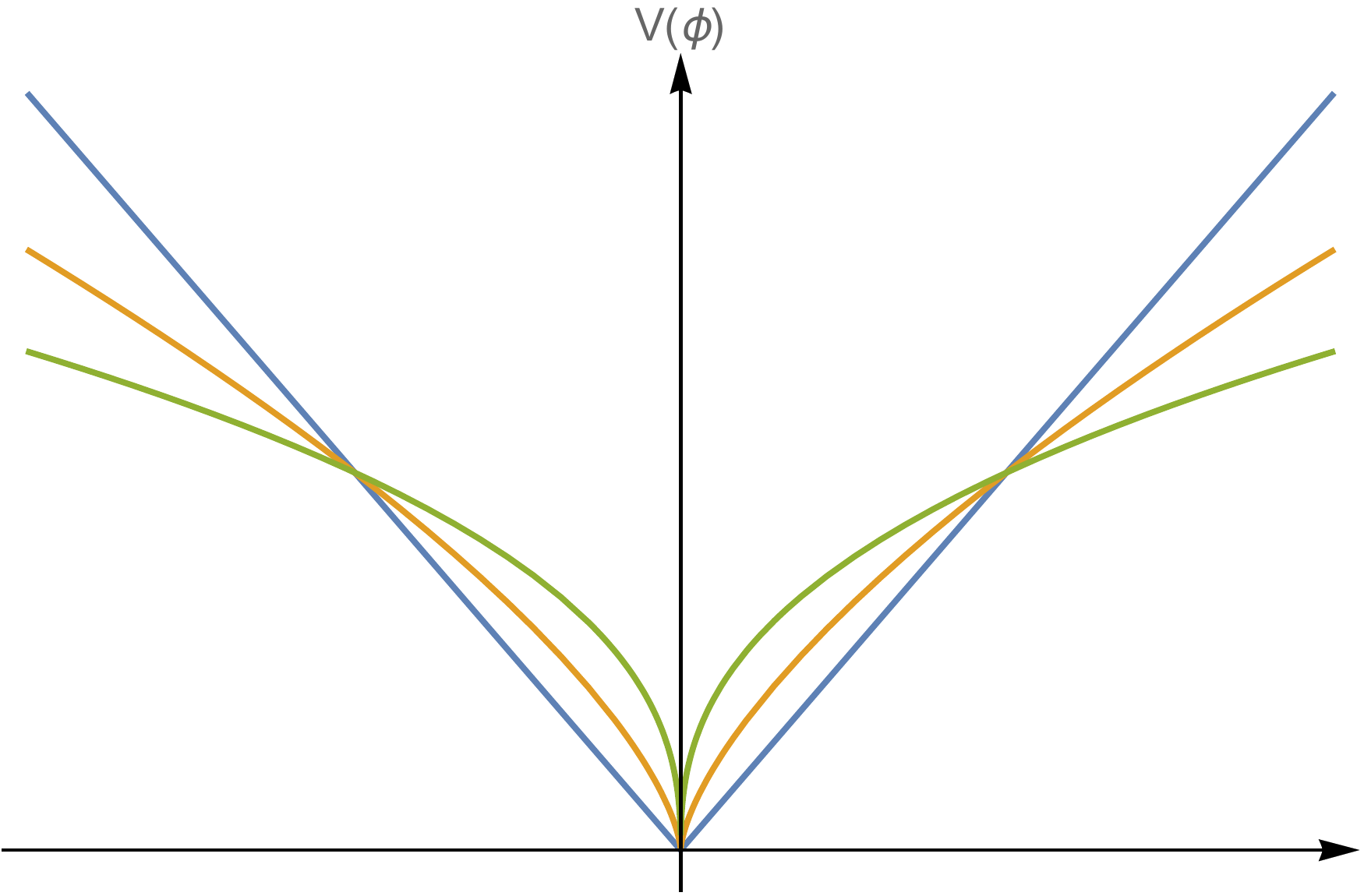}
\includegraphics[width=2.5in]{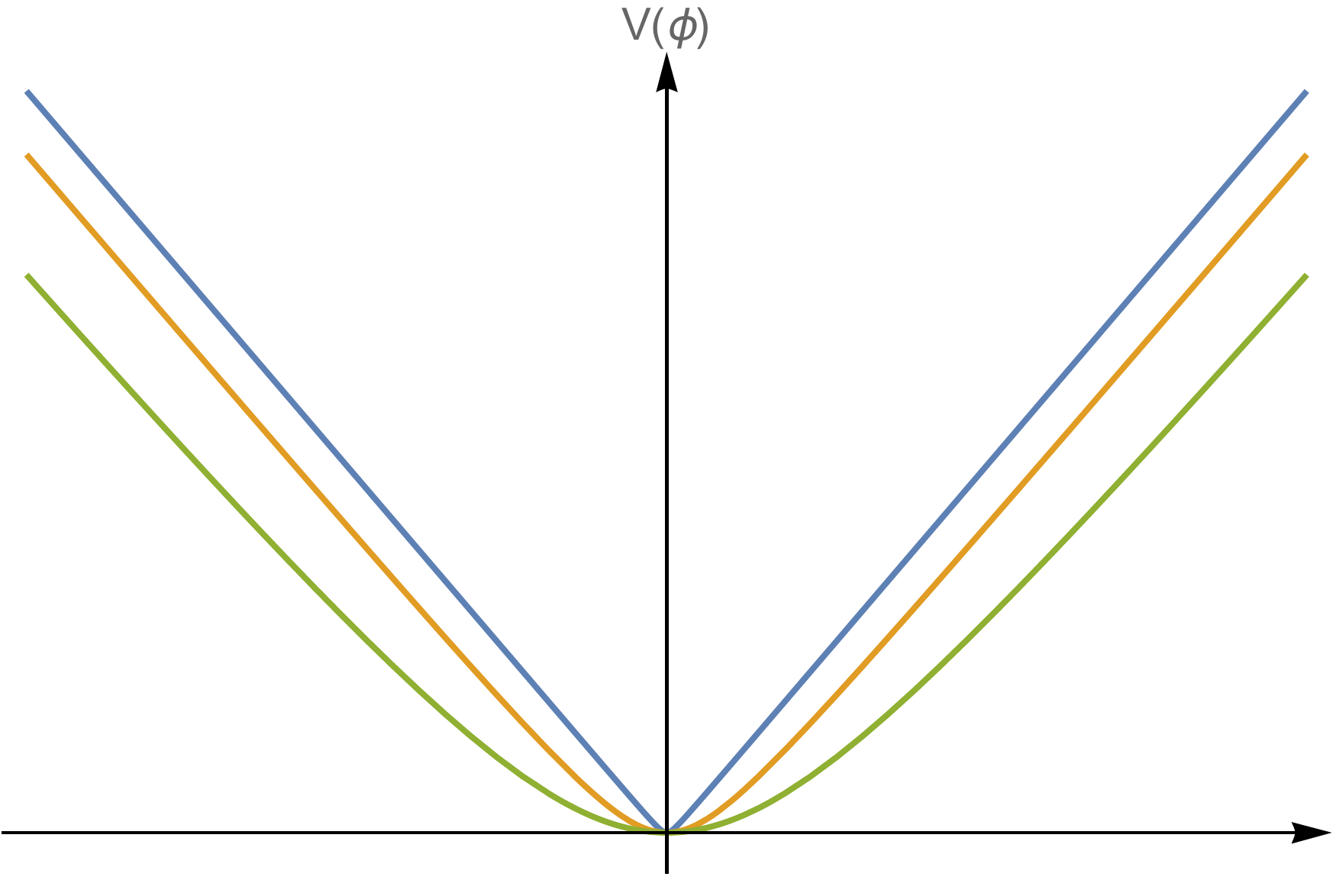}
\caption{Potentials~\eqref{eq:potentials} with $p=1$ (blue), $p=2/3$ (orange), $p=2/5$ (green) in the top panel,
and the $p=1$ potentials~\eqref{eq:pot2} with $M=0.001$ (blue), $M=0.01$ (orange) and $M=0.03$ (green) in the bottom panel.
}
\label{fig:pot}
\end{figure}

To investigate the effect of the cuspiness of the potentials~\eqref{eq:potentials} on the production of GWs,
we turn to a more general form of potentials
\begin{equation}
 \label{eq:pot2}
 V(\phi)=\frac{m^{2}M^{2}}{p}\left[\left(1+\frac{\phi^{2}}{M^{2}}\right)^{p/2}-1\right]\,,
\end{equation}
with an extended parameter $M$.
When $\phi/M$ is large, the potentials can be approximated by~\eqref{eq:potentials},
while when $\phi/M$ is small, the potentials become smooth near the minimum (see Fig.\ref{fig:pot}).

In a Friedmann-Robertson-Walker Universe the Friedman equation and the equation of motion of the scalar field are
\be
&& H^2 = \frac{1}{3M^2_\mathrm{pl}} \left\langle \frac12\dot{\phi}^2 + \frac{1}{2a^2}(\nabla\phi)^2 +V(\phi) \right\rangle \,, \\
&& \ddot{\phi}-\frac{1}{a^{2}}\nabla^{2}\phi+3H\dot{\phi}+\frac{dV}{d\phi}=0\,,
\label{eq:motion}
\ee
where $\langle...\rangle$ denotes a spatial average over the volume,
overdots denote derivatives with respect to the cosmic time $t$,
$H \equiv \dot{a}/a$ is the Hubble parameter and $\nabla$ is the spatial gradient.
GWs are described by the transverse-traceless gauge-invariant tensor perturbations $h_{ij}$,
i.e., $h^i_i=0$ and $h^i_{j,i}=0$.
The equation of motion of $h_{ij}$ is given by the linearized Einstein equation
\begin{equation}
\label{eq:EoMhij}
\ddot{h}_{ij}+3H\dot{h}_{ij}-\frac{1}{a^{2}}\nabla^{2}h_{ij}=\frac{2}{M_\mathrm{pl}^{2}a^{2}} T_{ij}^{\mathrm{TT}}\,,
\end{equation}
where $T^\mathrm{TT}_{ij}$ is the transverse-traceless (TT) projection of the anisotropic stress tensor $T_{ij}$.
In this paper we assume the inflaton is weakly coupled to other fields during preheating.
Actually GWs are sourced mainly by the inflaton fluctuations,
even if a parametric resonance for a matter field $\chi$ occurs in this model.
Although broad parametric resonance leads effectively to a fast growth of the fluctuations of $\chi$
if the inflaton is coupled to the field $\chi$,
our numerical simulations confirm that the growth of the inflaton fluctuations themselves triggered by the cusp in its potential
is more effective than that of the field $\chi$ by parametric resonance.
The anisotropic stress tensor takes the form
\begin{equation}
\label{eq:Tij}
T_{ij}=\partial_{i}\phi\partial_{j}\phi-\frac{1}{3}\delta_{ij}\partial^{k}\phi\partial_{k}\phi\,.
\end{equation}
The energy density of GWs is given by
\begin{equation}
\label{eq:EoMhij}
\rho_{\mathrm{GW}}=\frac{M^2_{\mathrm{pl}}}{4}\langle\dot{h}_{ij}\dot{h}^{ij}\rangle\,.
\end{equation}
It is convenient to introduce the dimensionless energy spectrum of GWs,
which is defined by
\be
\label{eq:es}
\Omega_\mathrm{GW}\equiv \frac{1}{\rho_c}\frac{d\rho_\mathrm{GW}}{d\ln k},
\ee
where $\rho_c \equiv 3M^2_\mathrm{pl}H^2$ is the critical density of the Universe.

\section{Numerical Algorithms}
\label{sec:nummet}
To obtain the energy spectrum of GWs produced during preheating,
one needs to solve numerically the equations of motion
of interacting scalar fields and tensor perturbations in an expanding Universe.
A number of codes have been developed to calculate the signal of GWs, including
the hybrid method, configuration-space method, Green's function method and pseudo-spectral method.
\begin{itemize}
\item Hybrid method. The scalar field equations~\eqref{eq:motion} are solved in configuration space
      while the GW equations~\eqref{eq:EoMhij} are solved in Fourier space by using fourth order
      Runge-Kutta integrator~\cite{Easther:2007vj}.
\item Configuration-space method. One first evolves both the tensor perturbations and scalar fields in configuration space,
      and then applies the transverse-traceless projector to the real physical $h_{ij}$ in Fourier space~\cite{GarciaBellido:2007af}.
\item Green's function method. The Green's function for the tensor perturbations $h_{ij}$ is constructed
      in Fourier space to directly calculate the energy spectrum of GWs~\cite{Dufaux:2007pt}.
      It is assumed that the modes of tensor perturbations are well inside the Hubble horizon.
\item Pseudo-spectral method. Both the tensor perturbations and scalar fields are evolved
      in Fourier space by using sencond/fourth order Runge-Kutta integration scheme~\cite{Zhou:2013tsa}.
      The nonlinear terms in the potential and its derivatives are computed by first converting the fields into configuration space,
      and then taking the inverse transform back to Fourier space.
\end{itemize}

The first three methods are based on LATTICEEASY~\cite{Felder:2000hq} that uses the finite-difference
method to compute spatial derivatives of the scalar fields
and the staggered leapfrog algorithm to compute time derivatives,
while the last one is based on PSpectRe~\cite{Easther:2010qz}
that uses the Fourier-space pseudo-spectral method to evolve the scalar fields.
In our simulations we adopt the configuration-space method for solving
the following evolution equation of the tensor perturbations in configuration space
\be
\label{eq:EoMuij}
\ddot{u}_{ij}+3H\dot{u}_{ij}-\frac{1}{a^{2}}\nabla^{2}u_{ij}=\frac{2}{M_\mathrm{pl}^{2}a^{2}}T_{ij}\,.
\ee
The source term is treated as an interaction term of $h_{ij}$ with the scalar field.
To avoid calculating the $\mathrm{TT}$ components at every step of the simulations,
we have defined a new quantity $u_{ij}$ in~\eqref{eq:EoMuij}.
Thus, the TT tensor perturbations can be written as
\be
h_{ij}(t,{\bf k})=\Lambda_{ij,lm}(\hat{\bf k})u_{lm}(t,{\bf k}),
\ee
where the TT projection operator $\Lambda_{ij,lm}(\hat{\bf k})$ is defined by
\be
\Lambda_{ij,lm}(\hat{\bf k}) \equiv P_{il}(\hat{\bf k})P_{jm}(\hat{\bf k}) - \frac12 P_{ij}(\hat{\bf k})P_{lm}(\hat{\bf k}),
\ee
with $P_{ij} \equiv \delta_{ij}-\hat{k}_i \hat{k}_j$, and
$u_{lm}(t,{\bf k})$ is the Fourier transform of the solution to Eq.~\eqref{eq:EoMuij}.
Similar to the equation of motion of the scalar field, we thus can evolve Eq.~\eqref{eq:EoMuij}
in configuration space using LATTICEEASY and
obtain $h_{ij}$ in Fourier space at any moment of the evolution by Fourier transform and TT projection.
In terms of $u_{ij}$, the energy density of GWs~\eqref{eq:EoMhij} can be expressed as~\cite{GarciaBellido:2007af}
\be
\rho_{\mathrm{GW}}
 =\frac{M_{\mathrm{pl}}^{2}}{4L^3}\int d^{3}\mathbf{k}\, \Lambda_{ij,lm}(\hat{\mathbf{k}}) \dot{u}_{ij}(t,\mathbf{k})\dot{u}_{lm}^{*} (t,\mathbf{k})\,.
\ee
Then the energy spectrum of GWs~\eqref{eq:es} becomes
\be
\Omega_{\mathrm{GW}}
 =\frac{M_{\mathrm{pl}}^{2}k^3}{4L^3\rho_c}\int d \Omega\, \Lambda_{ij,lm}(\hat{\mathbf{k}}) \dot{u}_{ij}(t,\mathbf{k})\dot{u}_{lm}^{*} (t,\mathbf{k})\,.
\ee

For this work we use a GPU-accelerated code based on OPENACC,
a performance-portable parallel programming model designed for scientists and engineers.
The expansion rate of the Universe is calculated self-consistently from spatially averaged energy density.
When the homogeneous field modes fragment into higher momentum modes,
second-order effects cannot be neglected and
thus the evolution of the nonlinear interaction has to be solved by lattice simulations.
We perform three-dimensional lattice simulations with $256^{3}$ points in a box with periodic boundary conditions.
The size of the box $L$ and the number of grid points per edge $N$ are in principle chosen
according to the physical features of the model.
In our simulations, the box size is chosen to be the resonance wavelength,
which is smaller than the Hubble horizon,
so that the interesting wavelengths such as the physical peaks in $\Omega_{\mathrm{GW}}$ are
located comfortably in between
the largest wavelength $L$ and the smallest wavelength $L/N$.

Initial conditions need to be set for the lattice calculations.
We set the initial values of the field as $\phi_i=1M_\mathrm{pl}$ for $p=1$
and $\phi_i = 0.4M_\mathrm{pl}$ for both $p=2/3$ and $2/5$.
Their derivatives determined by the inflation attractor are given by
$\dot{\phi}_i=-0.5M_\mathrm{pl}^{2}$ for both $p=1$ and $2/3$,
and $\dot{\phi}_i=-0.35M_\mathrm{pl}^{2}$ for $p=2/5$.
The initial values of the field fluctuations $\delta\phi_i$ and field derivative fluctuations $\delta\dot{\phi}_i$
are obtained from quantum vacuum fluctuations~\cite{Polarski:1995jg}
while the tensor perturbations and their derivatives are initialized as zero.
We set the scale factor $a_i = 1$ at the initial moment of the simulation.
In our package, the rescaled variables are used to reduce numerical errors,
\be
\label{eq:propara}
\tilde{\phi} &=& \frac{\phi}{\phi_i}a^{6/(2+p)}\,,\\
d\tilde{t} &=& dt\sqrt{\lambda M_\mathrm{pl}^{4-p}}\phi_i^{p/2-1}a^{(6-3p)/(2+p)}\,,\\
\kappa &=& \frac{k}{\sqrt{\lambda M_\mathrm{pl}^{4-p}}\phi_i^{p/2-1}}\,.
\ee
For $p<2$, the amplitude of oscillations of $\phi$ decreases and its period decreases as the Universe expands,
while the amplitude of oscillations of $\tilde{\phi}$ is invariant in the linear stage,
which makes the program more stable.

\begin{figure*}[t]
\includegraphics[width=3in]{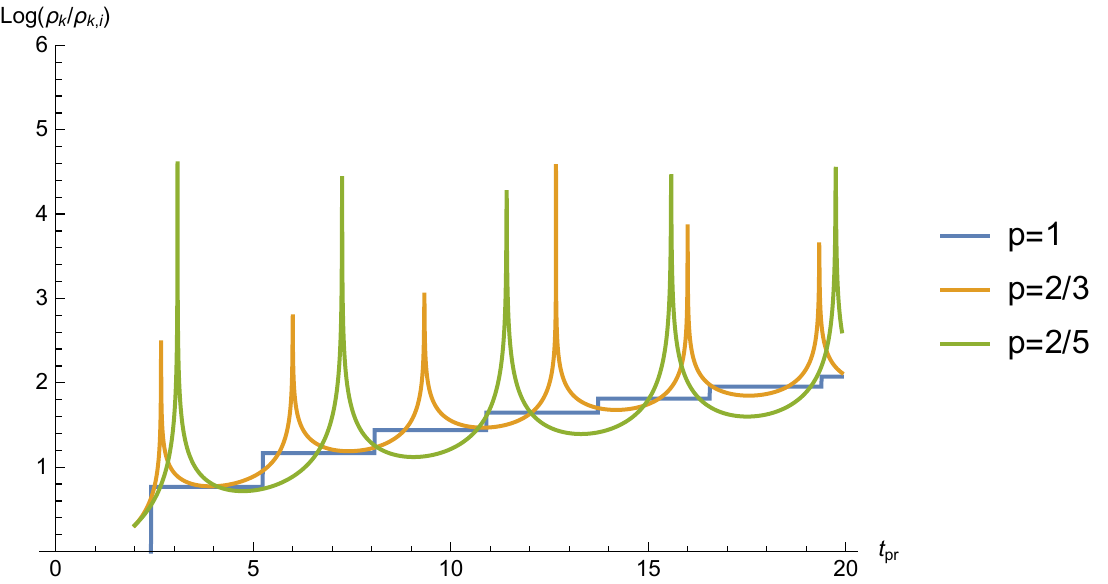}
\includegraphics[width=3in]{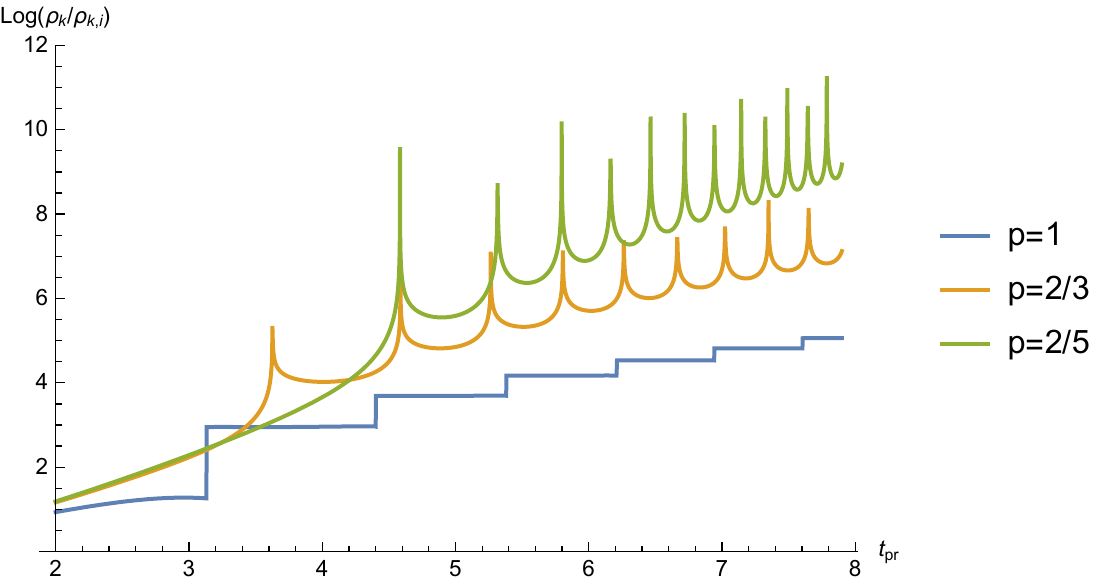}
\caption{
Evolutions of $\rho_k$ for the cuspy potentials~\eqref{eq:potentials} with $p=1$ (blue), $p=2/3$ (orange) and $p=2/5$ (green)
in Minkowski space (left panel) and an expanding Universe (right panel).
}
\label{fig:lin}
\end{figure*}

\section{Dynamics}
\label{sec:dyn}
\subsection{Linear Analysis}
\label{subsec:la}
For small fluctuations around a homogeneous field $\phi(t,\mathbf{x})=\bar{\phi}(t)+\delta \phi(t,\mathbf{x})$.
After inflation, the field begins to oscillate around the minimum of its potential.
In the linear analysis, each Fourier mode of fluctuations evolves independently
and thus the fluctuation equation can be numerically solved as an ordinary differential equation.
For the model~\eqref{eq:potentials} with $p<1$, which we shall refer to as the infinite
cuspy model, we have to deal with a diverging  $dV/d\bar{\phi}$ near $\bar{\phi}=0$.
To avoid the singularity of $dV/d\bar{\phi}$ at $\bar{\phi}=0$ in the linear analysis,
we set $M=0.001M_{\mathrm{pl}}$ in the model~\eqref{eq:pot2} to approximate the model~\eqref{eq:potentials}.
As we shall see, if $M$ is smaller than a threshold related to the initial conditions for $\bar{\phi}$,
the simulation results are independent of the value of $M$.
The rescaled time $t_{pr} \equiv tM_\mathrm{pl}/\sqrt{\lambda}$ and rescaled wavenumber $k/\sqrt{\lambda}M_\mathrm{pl}$ are used so that the period of field oscillations and wavenumber of resonant modes are $\mathcal{O}(1)$.
The equation of motion of the homogeneous field can be solved independently assuming field fluctuations have little effect on it.
Hence the Hubble parameter is calculated from the energy density of the homogeneous field
when the equation of motion of fluctuations is solved in an expanding Universe.
The linear approximation is valid
until the field oscillates $\mathcal{O}(10)$ times.
As an illustration, in linear analysis we choose $\bar{\phi}_i=1M_\mathrm{pl}$ and $\dot{\bar{\phi}}_i=0$
as the initial conditions for the homogeneous field.
The evolution of fluctuations is plotted in Fig.~\ref{fig:lin} in terms of $\rho_k/\rho_{k,i}$
so that these results are independent of the initial conditions for $\delta\phi$ and $\delta\dot{\phi}$.
Here the energy density spectrum of the field is defined as
\be
\label{eq:rhok}
k^3 \rho_{k}=\dfrac{1}{2} k^3 \left[|\partial_{\tau}(a\delta\phi_{k})|^{2}+\omega^2_{k}|a\delta\phi_{k}|^{2}\right]\,,
\ee
where $\omega_{k}^{2}=k^{2}+a^{2}\langle V_{\phi\phi}\rangle-\partial^2_\tau a / a$,
$\delta \phi_k$ is the Fourier modes of $\delta \phi$ and $\tau$ is the conformal time.

We now begin with the linear analysis for the evolution of fluctuations.
When $\bar{\phi}$ approaches the minimum of the potential, the effective mass changes rapidly and the adiabatic approximation becomes invalid.
The non-adiabatic production of particles occurs only near $\phi=0$ in the $p=1$ case
while the tachyonic growth of fluctuations persists continuously in the $p< 1$ case.
The equation of motion of $\delta\phi$ in Fourier space reads
\begin{equation}
\label{eq:EOMdelta}
\ddot{\delta\phi_{k}}+\frac{k^{2}}{a^{2}}\delta\phi_{k}+3H\dot{\delta\phi_{k}}+\frac{d^2V}{d\bar{\phi}^2}\delta\phi_{k}=0\,.
\end{equation}
The oscillations of $\bar{\phi}$ are periodic if the expansion of the Universe is neglected.
According to the Floquet theory, Eq.~\eqref{eq:EOMdelta} has a general solution
\begin{equation}
\label{eq:EOMFlo}
\delta\phi_{k}=\mathcal{P}_{k+}(t)\exp(\mu_{k}t)+\mathcal{P}_{k-}(t)\exp(-\mu_{k}t)\,,
\end{equation}
where $\mathcal{P}_{k\pm}$ are periodic functions that are determined by the initial conditions
and $\mu_{k}$ are the Floquet exponents.
If the real part of $\mu_{k}$ is nonzero, i.e., $\mathrm{Re}(\mu_{k}) \neq 0$,
$\delta\phi_{k}$ is unstable and fluctuations grow exponentially.

We now focus on the $p=1$ case.
As $\bar{\phi}$ crosses the minimum of the potential,
a finite sudden change of $dV/d\bar{\phi}$ takes place.
Thus this case is referred to as the finite cuspy model.
In the $p<1$ case, the difference between $dV/d\bar{\phi}$ of two branches diverges, so we call it the infinite cuspy model,
which will be discussed later.
Integrating Eq.~\eqref{eq:EOMdelta} over an infinitesimal integration interval containing the time when $\bar{\phi}$ crosses the minimum,
we find a sudden change of $\delta\dot{\phi}_k$ takes place.
To understand the evolution of fluctuations in the infrared region,
it is convenient to consider the case in which the period of $\delta\phi_k$ is much longer than that of $\bar{\phi}$.
We neglect the expansion of the Universe for the moment and thus the friction term drops out of the equation of motion.
Assuming $k=0$ Eq.~\eqref{eq:EOMdelta} is simplified to
\begin{equation}
\label{eq:EOMdelta2}
\delta \ddot{\phi}+\frac{d^2V}{d\bar{\phi}^2}\delta\phi=0\,,
\end{equation}
In the $p=1$ case the derivative of the potential with respect to $\bar{\phi}$ is a step function
and its second order derivative is a delta function.
Since $\dot{\bar{\phi}}$ can be approximated by the maximum value of $\dot{\bar{\phi}}$ in a small vicinity of $\bar{\phi}=0$,
the sudden change of $\delta\phi$ is quantified as
\begin{equation}
\begin{split}
\label{eq:deltaphigap}
\Delta\delta\dot{\phi} = -K\delta\phi\,,\\
K\equiv \frac{2\lambda M_\mathrm{pl}^{3}}{|\dot{\bar{\phi}}_{m}|}\,,
\end{split}
\end{equation}
where $\dot{\bar{\phi}}_m$ is the maximum value of $\dot{\bar{\phi}}$ at $\bar{\phi}=0$.
$\delta\phi_{j}$ represents the value of $\delta\phi$ shortly before the $j$th sudden change.
Since $\delta\dot{\phi}$ is invariant between two adjacent sudden changes,
the relationships between $\delta\phi_{j+1}$, $\delta\dot{\phi}_{j+1}$, $\delta\phi_{j}$ and $\delta\dot{\phi}_{j}$ are
\begin{equation}
\label{eq:recurrence}
\delta\phi_{j+1}=\delta\phi_{j}+T\delta\dot{\phi}_{j+1}\,, \quad \delta\dot{\phi}_{j+1}=\delta\dot{\phi}_{j}-K \delta\phi_{j}\,,
\end{equation}
where $T$ denotes the time interval between two adjacent sudden changes.
We have $T=|2\dot{\bar{\phi}}_{m}|/\lambda M_\mathrm{pl}^{3}$, and $K =4/T$.
The general solution of $\delta\phi_{j}$ is
\begin{equation}
\label{eq:soluA}
\delta\phi_{j}=(-1)^{1+j}\left[j(2\delta\phi_{1} - T\delta\dot{\phi}_{1}) + T\delta\dot{\phi}_{1} - \delta\phi_{1}\right]\,,
\end{equation}
which indicates that $\delta\phi_{j}$ increases linearly with $j$
when $\bar{\phi}$ begins to oscillate around the minimum of its potential. 
Taking the expansion of the Universe into consideration,
the amplitude and period of $\bar{\phi}$ decreases due to the Hubble friction.
We find the amplitude is proportional to $t^{-2}$ and the period is proportional to $t^{-1}$.
The amplitude of fluctuations increases as $t^{2}$.
Nonzero modes of fluctuations will begin to grow exponentially soon.
From Fig.~\ref{fig:lin} we can see that the period of fluctuations decreases
in the expanding Universe.

\begin{figure*}[t]
\includegraphics[width=3in]{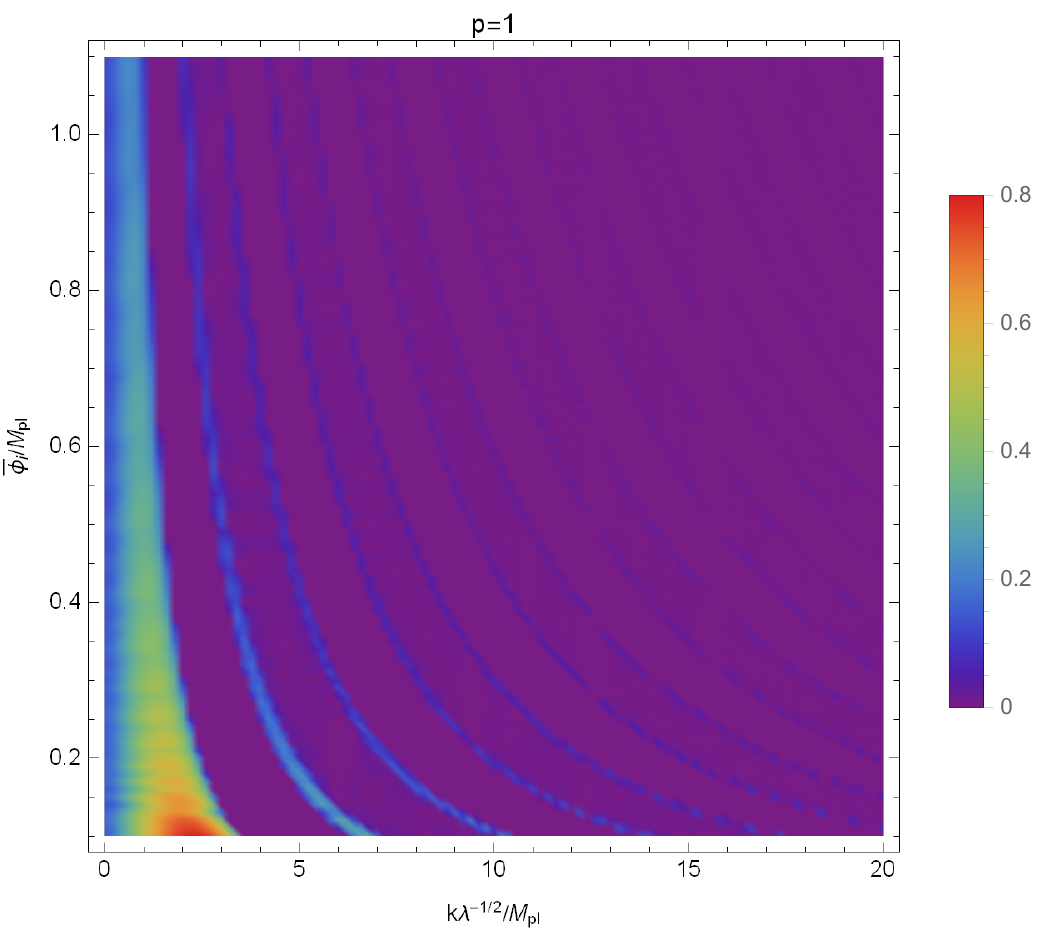}
\includegraphics[width=3in]{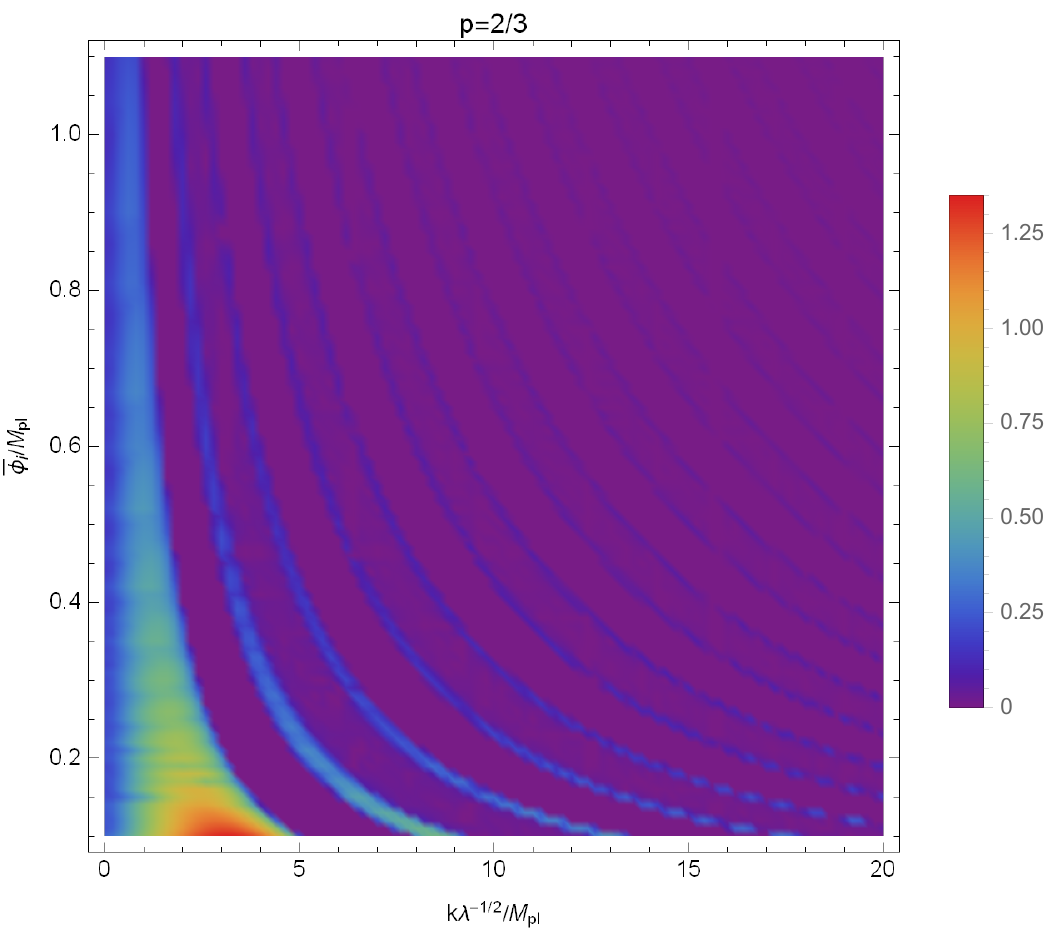}
\includegraphics[width=3in]{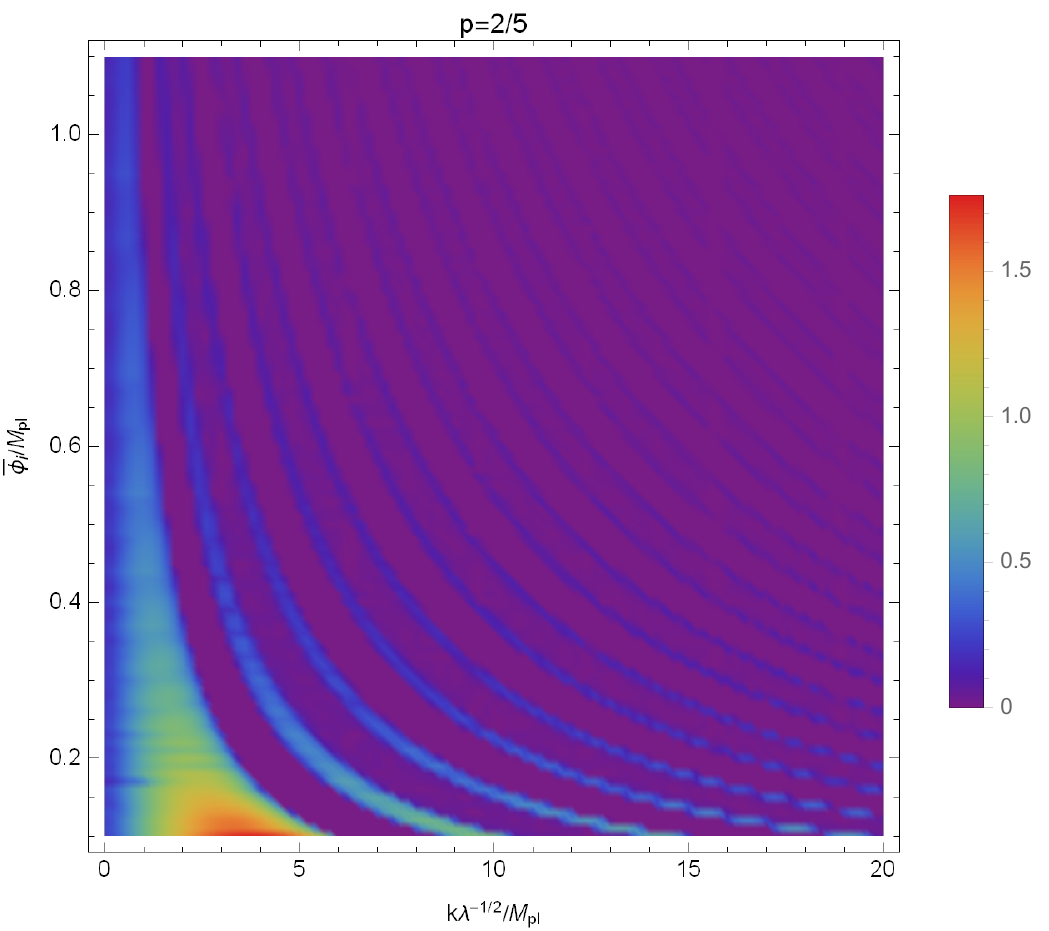}
\includegraphics[width=3in]{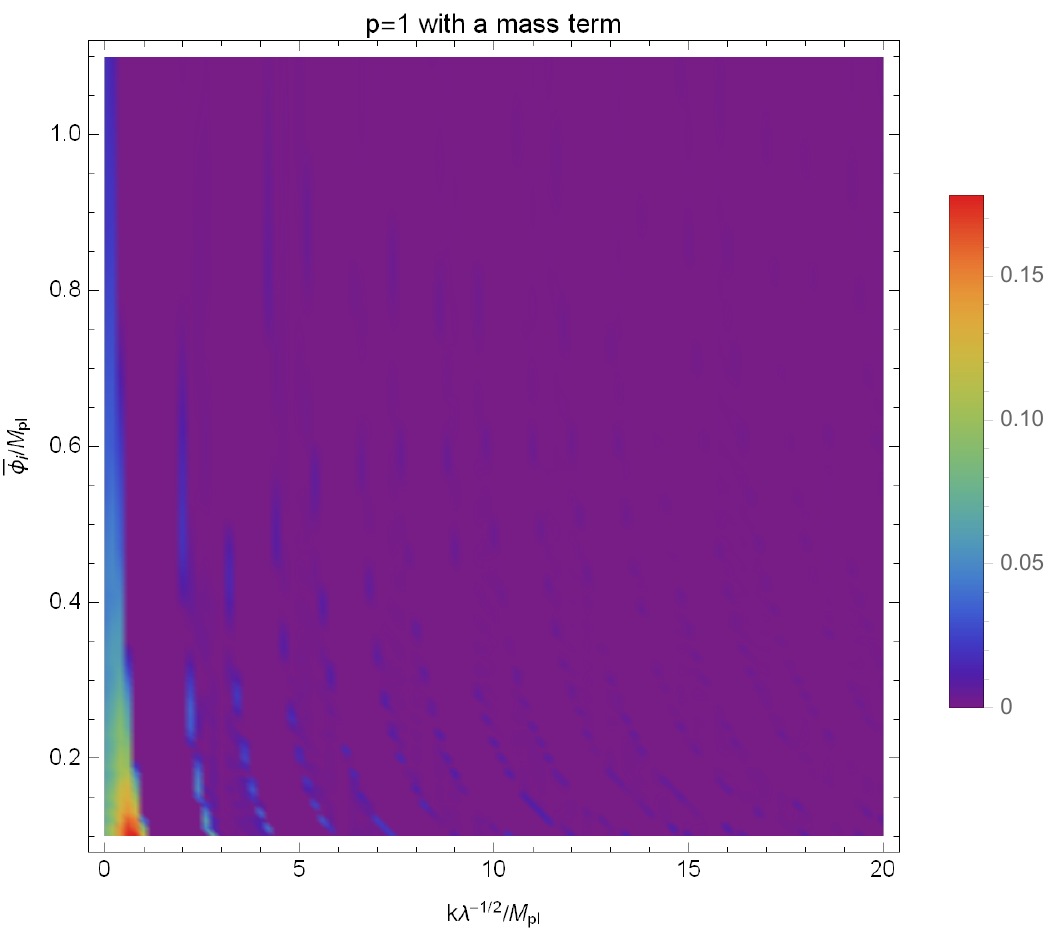}
\caption{
Resonance strength $|\mathrm{Re}(\mu_{k})|$ as a function of $k$
and $\bar{\phi}_i$ for the cuspy potentials~\eqref{eq:potentials} with $p=1$ (top-left panel),
$p=2/3$ (top-right panel), $p=2/5$ (bottom-left panel) and the potential~\eqref{eq:potmass} (bottom-right panel).
}
\label{fig:mu}
\end{figure*}

To investigate the evolution of nonzero modes of fluctuations, a semi-analytical method and numerical simulations are applied.
If the $k$ mode of fluctuations lies in the resonance band, $\mathrm{Re}(\mu_{k}) \neq 0$ means the mode grow exponentially.
We firstly estimate the value of $k_{m}$ corresponding to the maximum of $|\mathrm{Re}(\mu_{k})|$.
Intuitively, if the natural period of $\delta\phi$, which is the period in the absence of the source term in~\eqref{eq:EOMdelta},
coincides with the period of the source term, $\delta\phi$ will increase exponentially.
As expected, each sudden change of $\delta\dot{\phi}$ can effectively transfer energy into fluctuations.
From~\eqref{eq:deltaphigap} we see that the maximum of $\Delta\delta\dot{\phi}$
at the sudden change corresponds to the maximum of $\delta\phi$.
Moreover, the sign of $\delta\dot{\phi}$ does not change at the sudden change.
For the $k_m$ mode, when $\bar{\phi}$ reaches the minimum of the potential,
$|\delta\phi|$ reaches its maximum, $|\delta\phi_{m}|$, and $\delta\dot{\phi}=0$.
There are infinite $k$ modes which satisfy the condition.
Among them the smallest $k$ corresponds to the maximum of $|\mathrm{Re}(\mu_{k})|$
because the amplitude of $\delta\phi$ is larger for smaller $k$
with the same $\rho_k$, which is consistent with our numerical results in Fig.~\ref{fig:mu}.
The time interval between two adjacent sudden changes can be estimated
by a quarter of the natural period of $\delta\phi$, i.e., $\pi a/2k$.
In this case it is equal to half of the period of $\bar{\phi}$.
Thus $k_{m}$ reads
\begin{equation}
\label{eq:muk}
k_{m}=\frac{\lambda M_{\mathrm{pl}}^{3}\pi a}{4|\dot{\bar{\phi}}_{m}|}\,,
\end{equation}
where
\be
    |\dot{\bar{\phi}}_{m}|&=\frac{\sqrt{2\lambda\bar{\phi}_{i}M_{\mathrm{pl}}^{3}}}{a}\,.
\ee
The maximum of $|\mathrm{Re}(\mu_{k})|$ is
\begin{equation}
\label{eq:Ek}
    |\mathrm{Re}(\mu_{k})|_{m} = \frac{8a}{\pi^{2}}\sqrt{\frac{2\lambda M_{\mathrm{pl}}^{3}}{\bar{\phi}_i}}\,.
\end{equation}


The resonance strength $|\mathrm{Re}(\mu_{k})|$ in the $p=1$ case is plotted in the top-left panel of Fig.~\ref{fig:mu}.
We can see there are some resonance bands and the first band is the dominant one.
Since the physical momentum is redshifted by the expansion of the Universe,
the modes with $k> k_{m}$, which are stable at the beginning,
gradually enter the resonance band and ultimately exit the band.
As the Universe expands, the homogeneous field oscillates faster
and then the sudden changes take place more frequently.
Hence the modes of fluctuations, which enter the resonance band later,
have larger Floquet exponents and may exceed the modes in the first resonance band.

\begin{figure*}[t]
\includegraphics[width=2.2in]{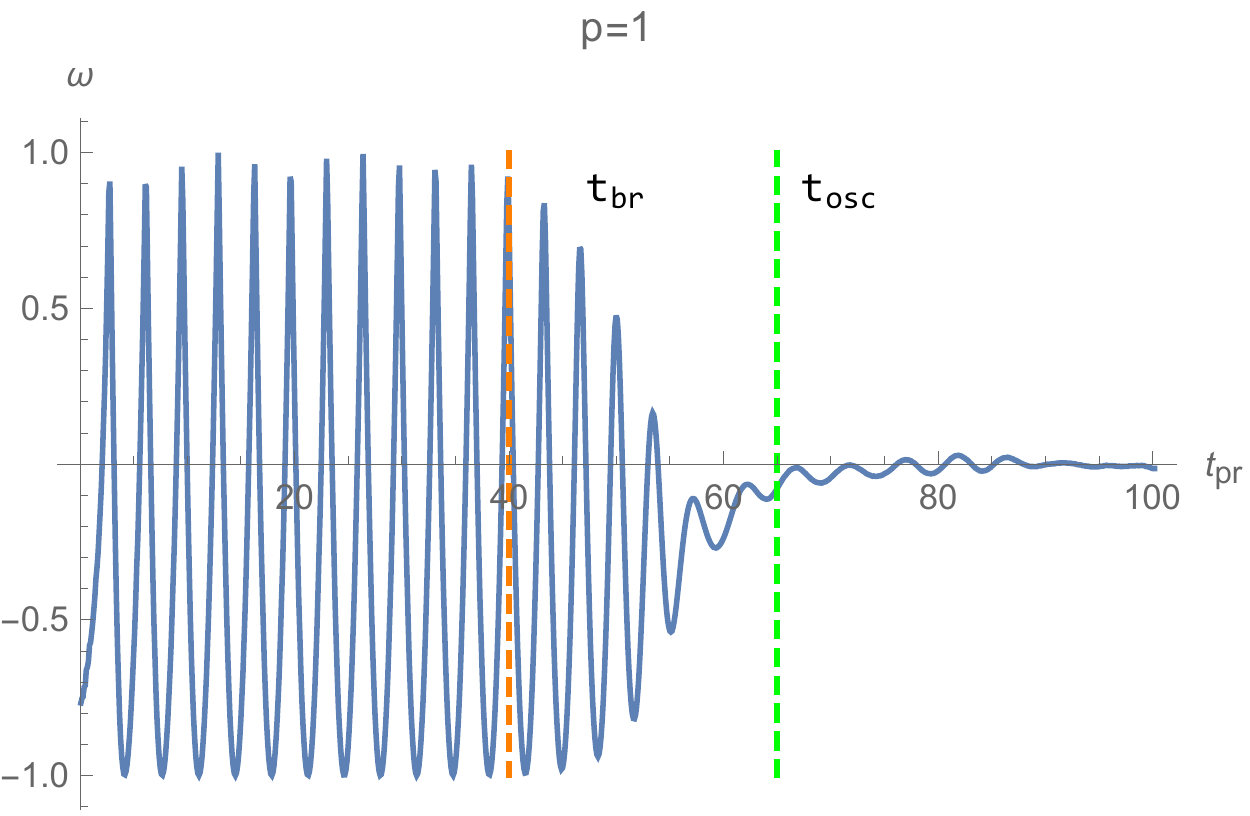}
\includegraphics[width=2.2in]{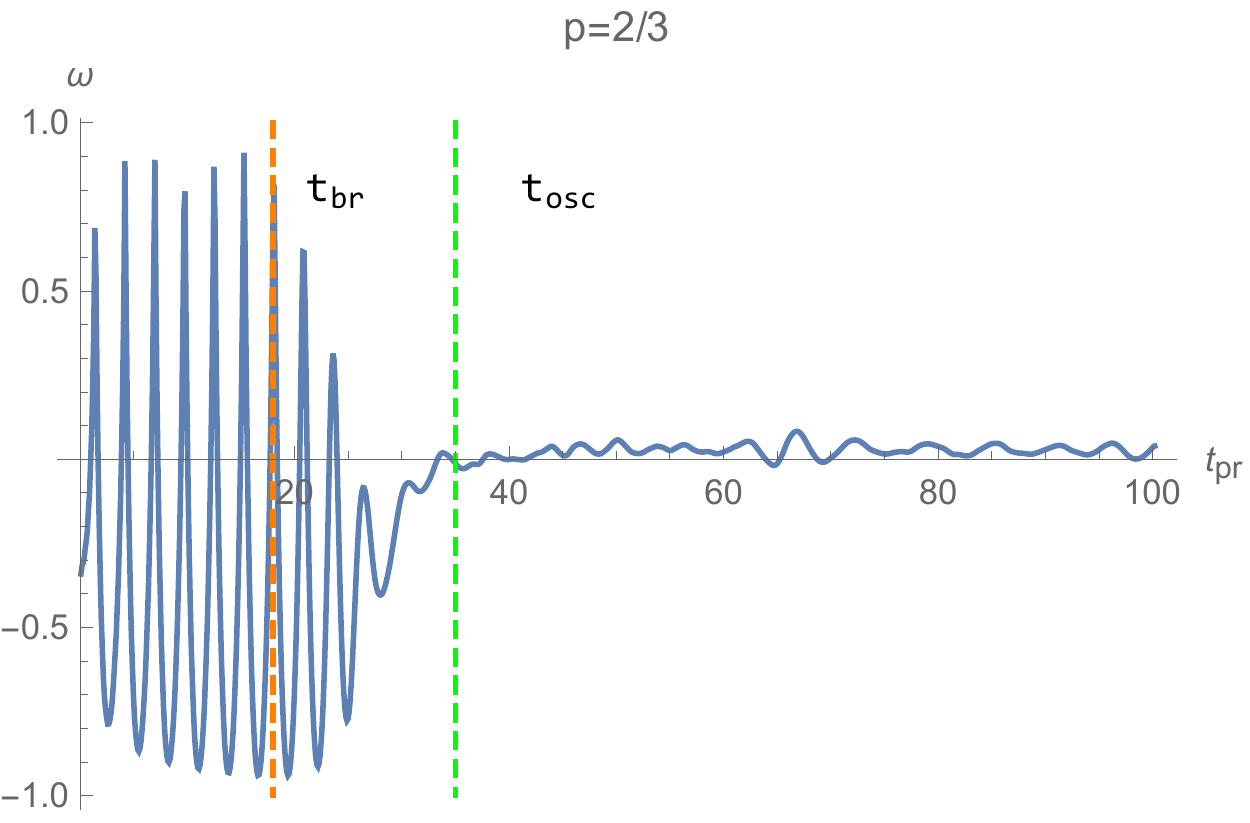}
\includegraphics[width=2.2in]{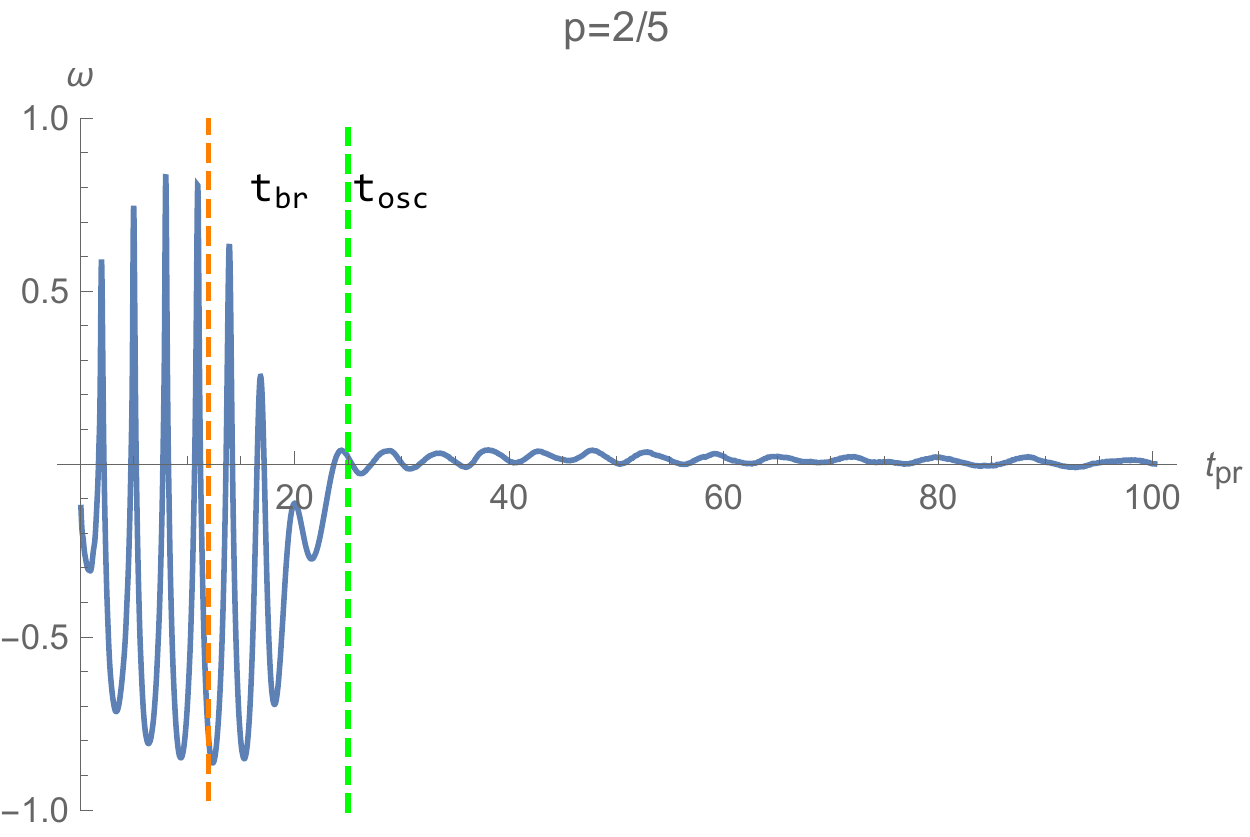}
\includegraphics[width=2.2in]{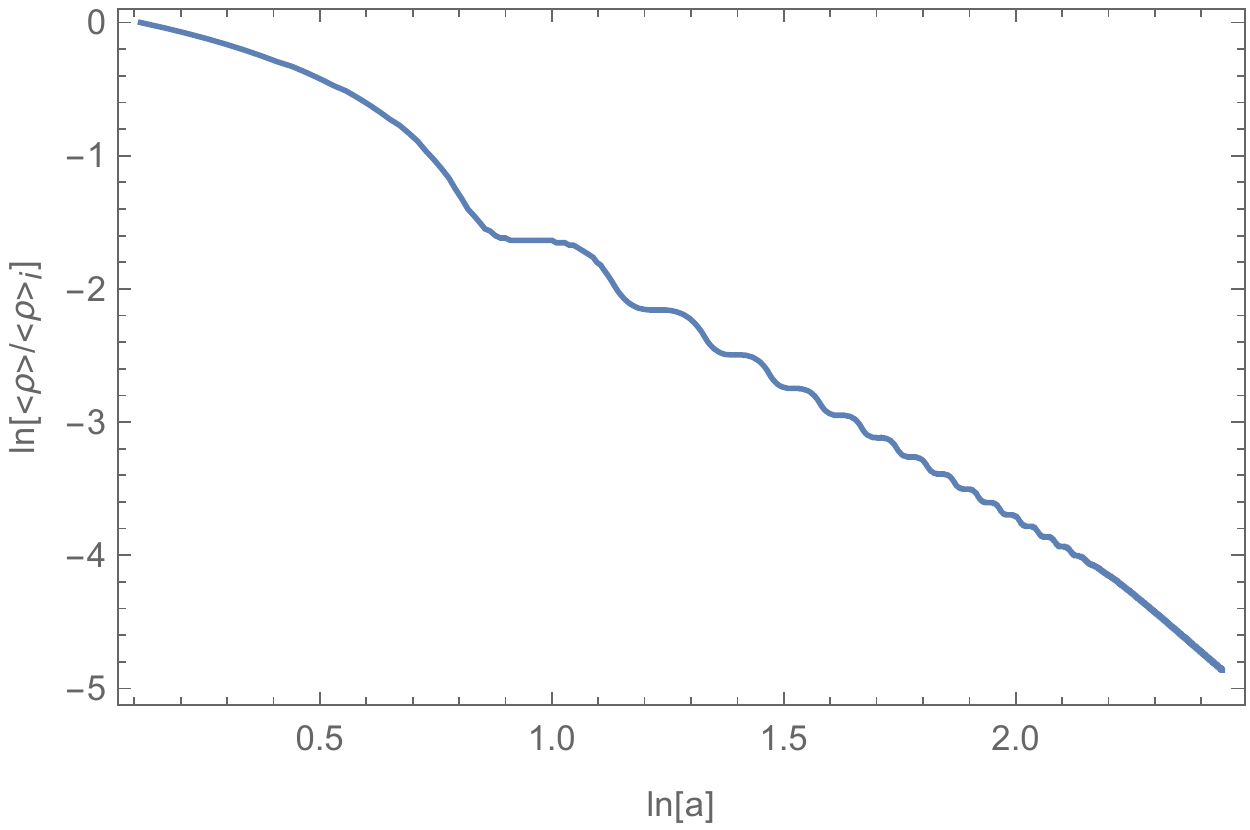}
\includegraphics[width=2.2in]{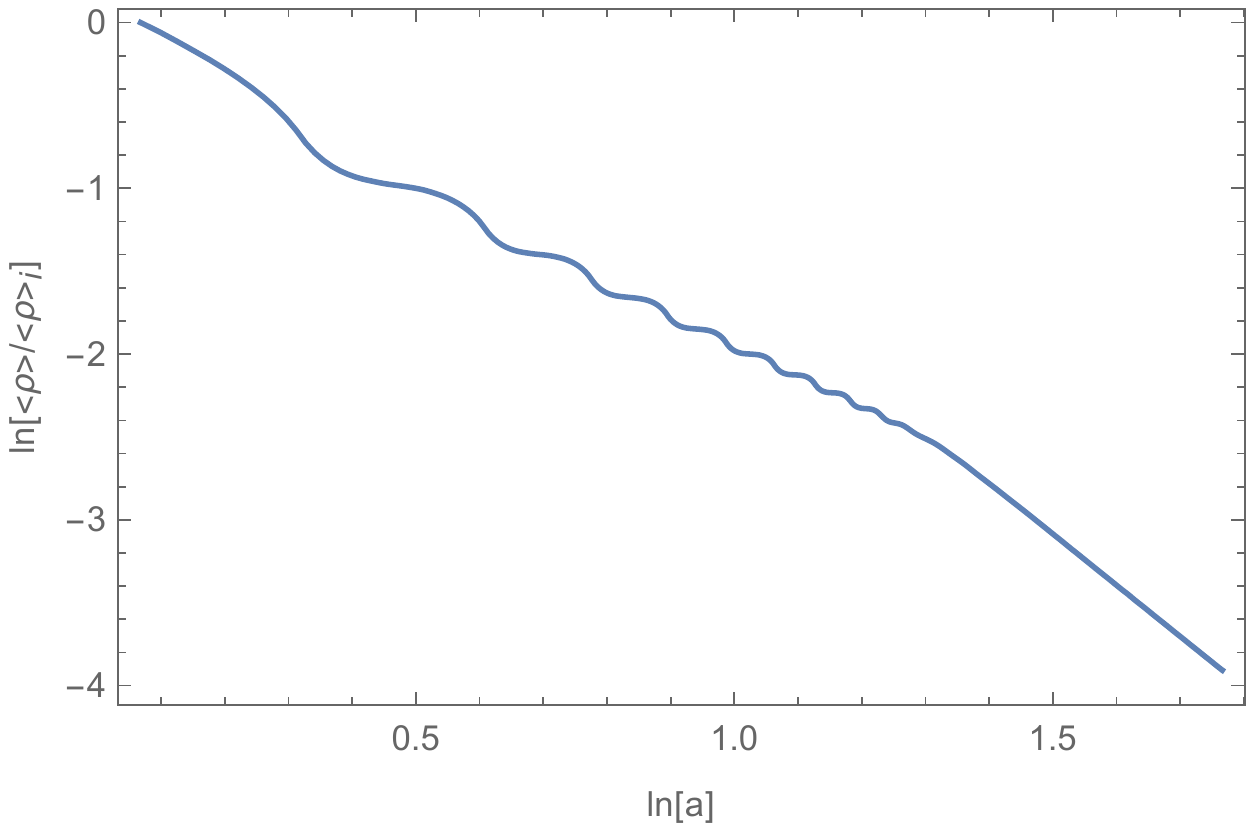}
\includegraphics[width=2.2in]{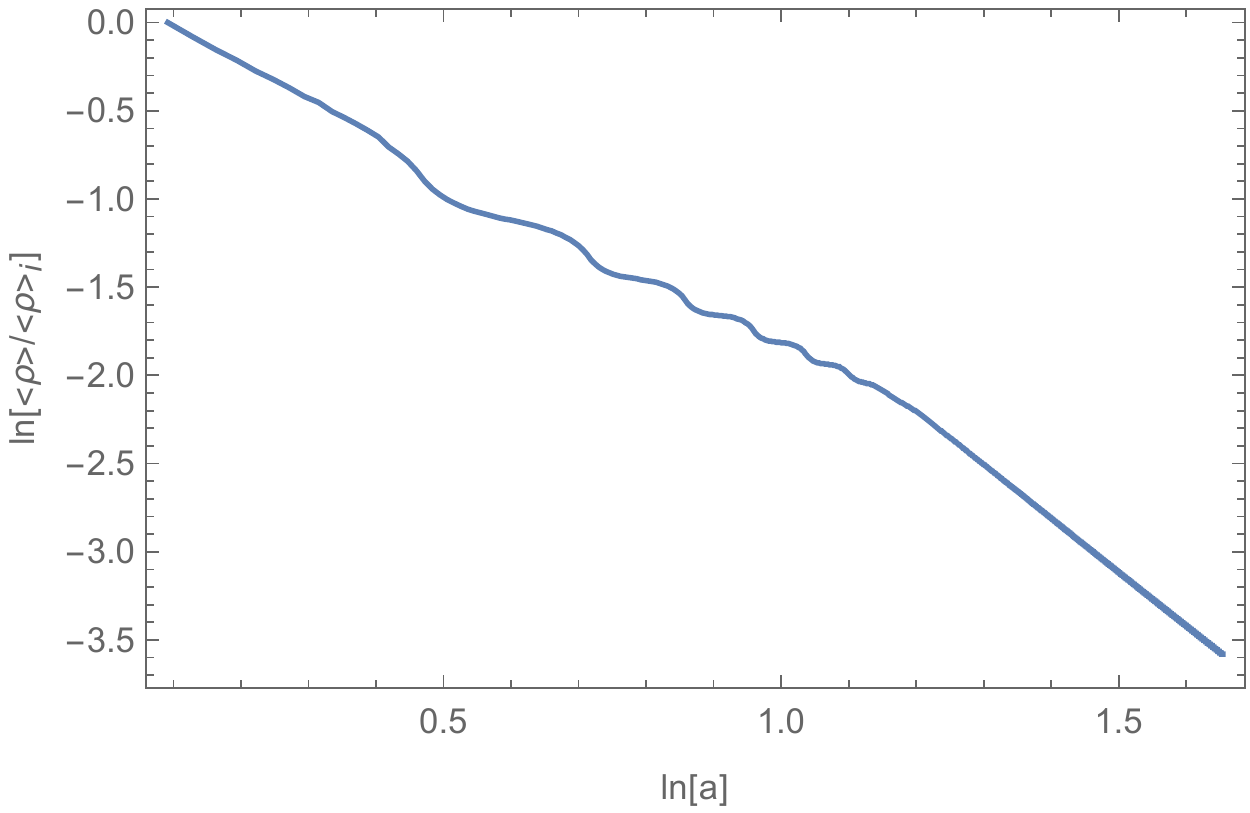}
\caption{
Evolutions of the EoS parameter $\omega$ for the cuspy potentials~\eqref{eq:potentials} with $p=1$ (top-left panel),
$p=2/3$ (top-middle panel) and $p=2/5$ (top-right panel),
and the evolutions of the corresponding energy density $\rho$ in the bottom panel.
}
\label{fig:omega}
\end{figure*}

As for the infinite cuspy model, using numerical simulations, we plot $|\mathrm{Re}(\mu_{k})|$
in the top-right and bottom-left panels of Fig.\ref{fig:mu}.
When $\bar{\phi}$ approaches $\bar{\phi}=0$, both $dV(\bar{\phi})/d\bar{\phi}$ and $\delta\dot{\phi}$ increase to infinity.
When $\bar{\phi}$ crosses $\bar{\phi}=0$, they suddenly change their signs
and subsequently their absolute values decreases.
This leads to the divergences of the energy density spectrum at the moment of $\bar{\phi}=0$, as shown in Fig.~\ref{fig:lin}.
Interestingly, the expansion of the Universe accelerates the growth of fluctuations
because the amplitude of $\bar{\phi}$ decreases due to the Hubble friction.

Actually, even if the potential contains an effective mass term
\begin{equation}
 \label{eq:potmass}
 V(\phi)=\lambda M_\mathrm{pl}^{3}|\phi| + \lambda_1 M_\mathrm{pl}^{4-2q}\phi^{2q}\,,
\end{equation}
our conclusions still hold.
For example, the bottom-right panel of Fig.~\ref{fig:mu} shows the resonance strength $|\mathrm{Re}(\mu_k)|$
in the model~\eqref{eq:potmass} with $q=1$.
Compared to the model~\eqref{eq:potentials} with $p=1$,
from Fig.~\ref{fig:mu} we can see that the resonance bands are shifted to small $k$
and become narrow due to the effective mass term.
It means the resonance can occur even for the zero mode of fluctuations.
Moreover, the resonance strength is suppressed in the presence of the effective mass term.
We find that if the potential is dominated by the effective mass term,
after several oscillations the linear term becomes dominant as $\bar{\phi}$ decreases.
Therefore, our analysis is still valid in the nonlinear stage.

\subsection{Nonlinear dynamics}
\begin{figure*}[t]
\includegraphics[width=2.2in]{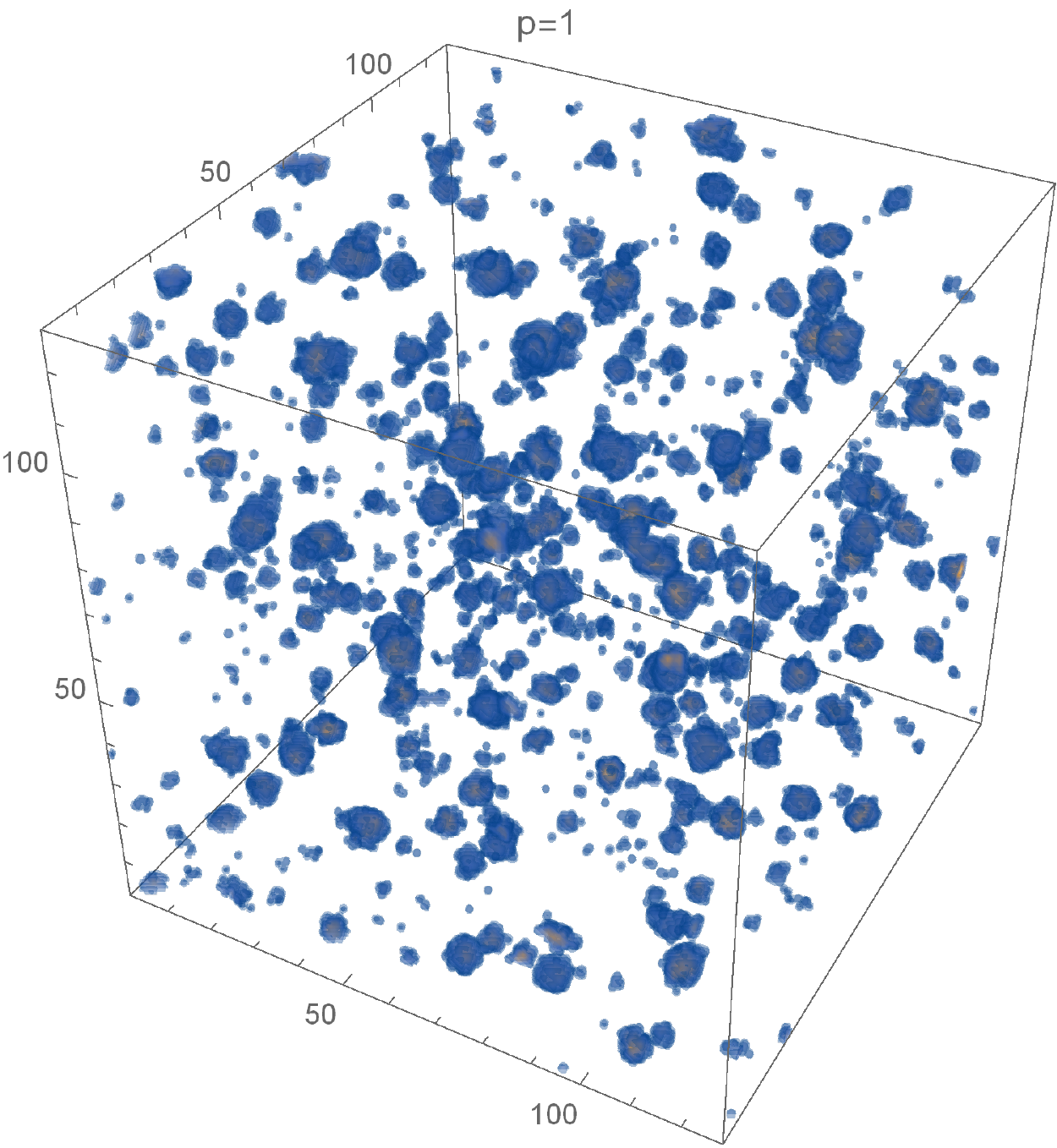}
\includegraphics[width=2.2in]{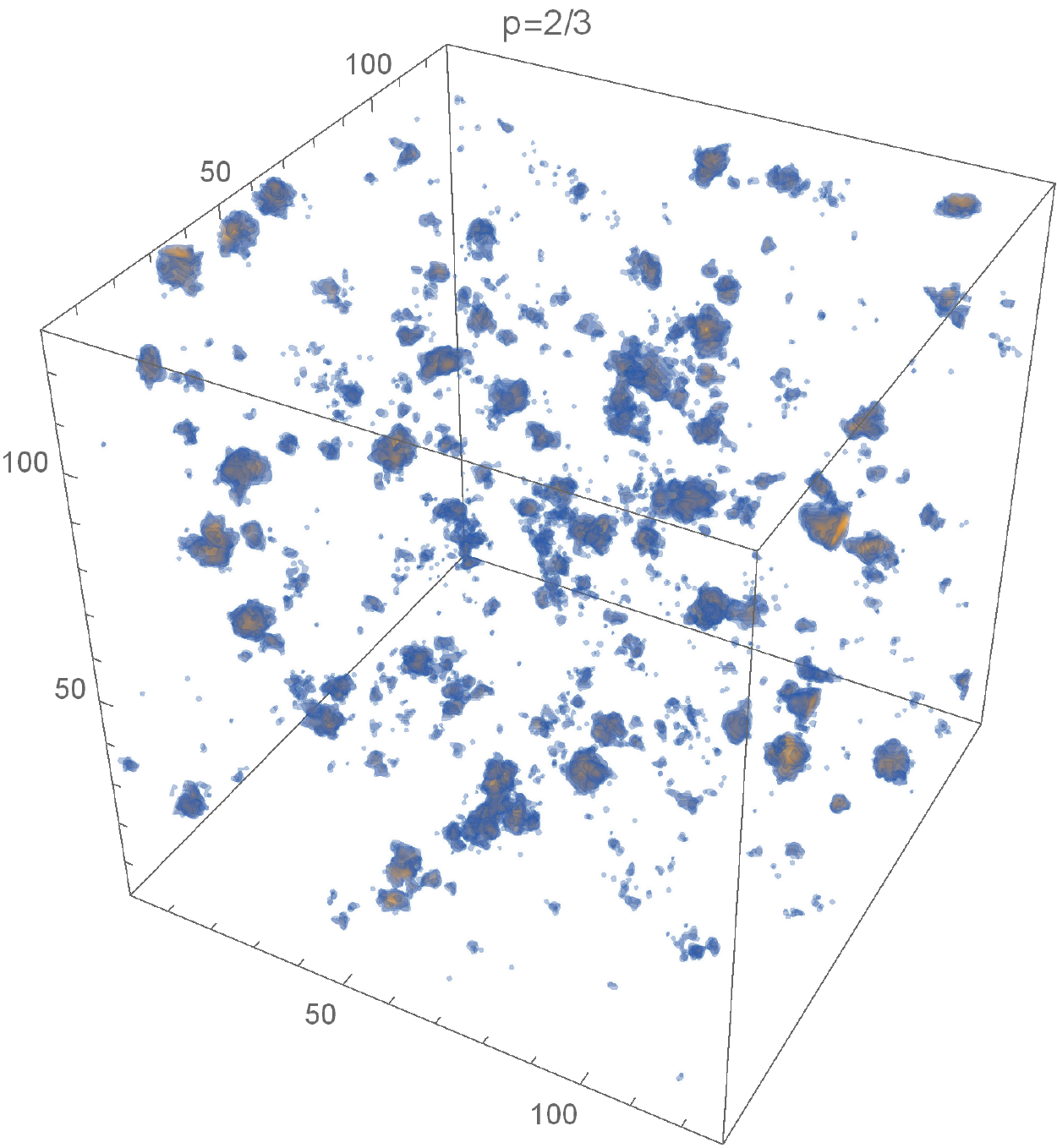}
\includegraphics[width=2.2in]{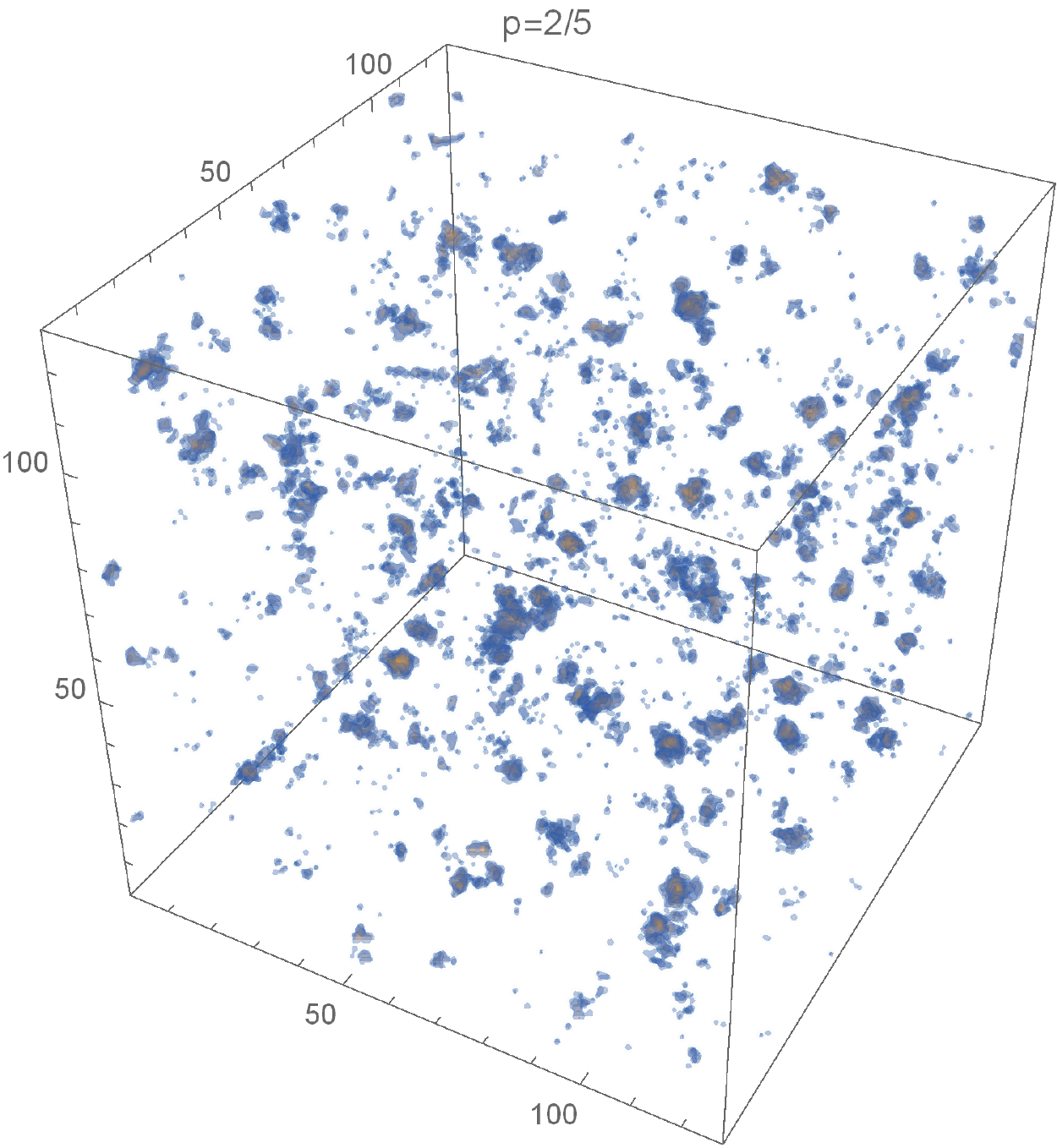}
\includegraphics[width=2.2in]{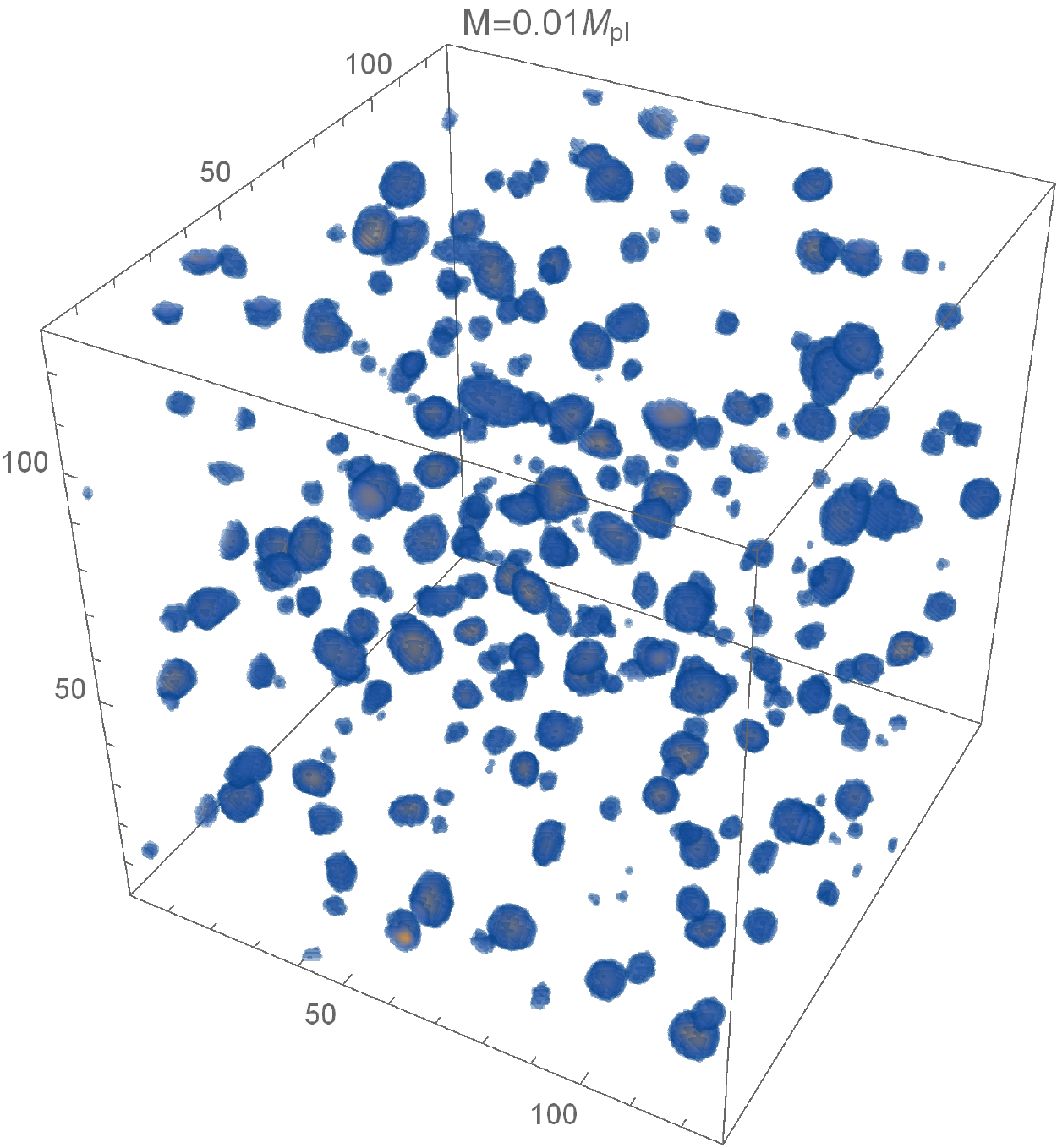}
\includegraphics[width=2.2in]{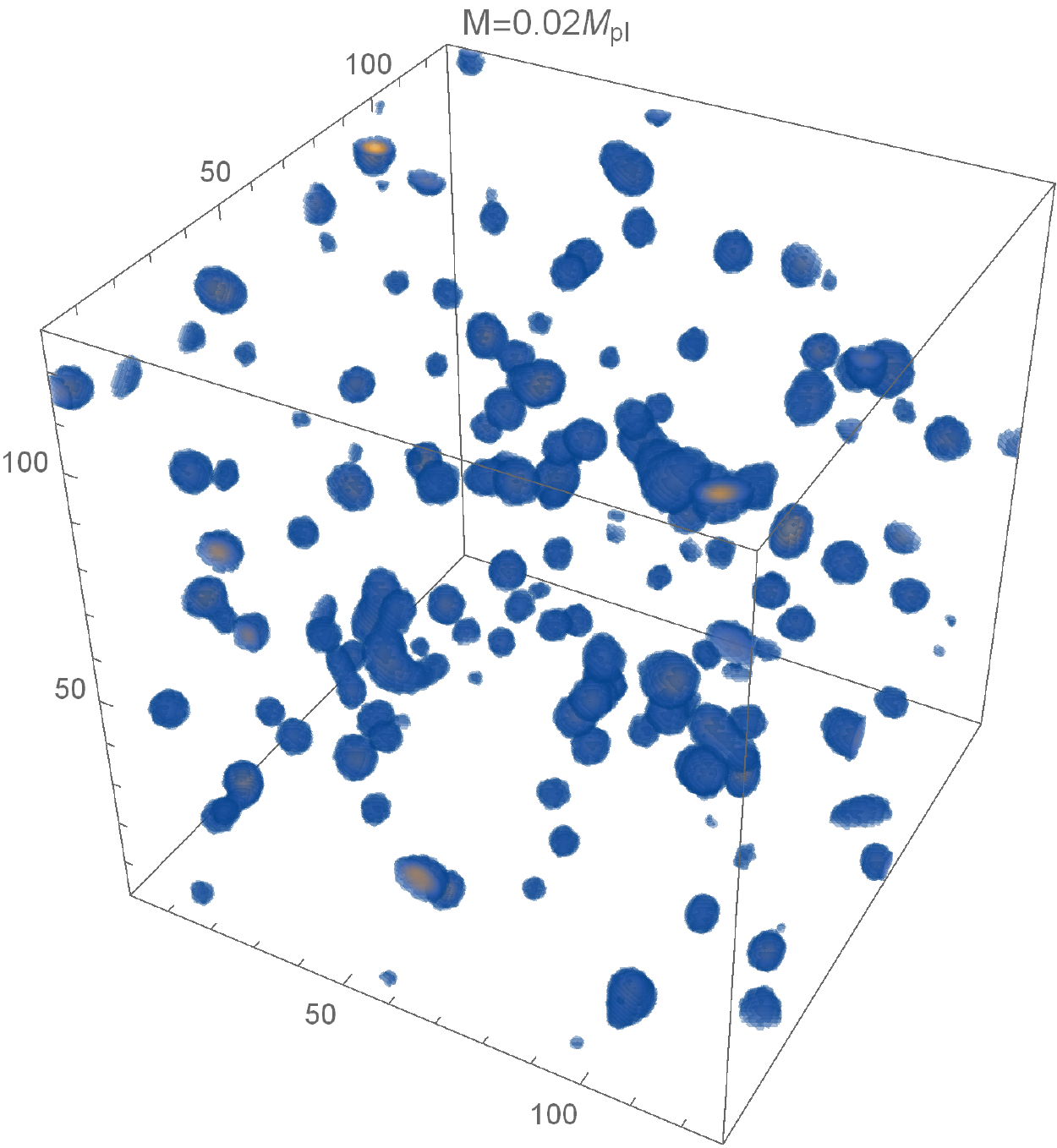}
\includegraphics[width=2.2in]{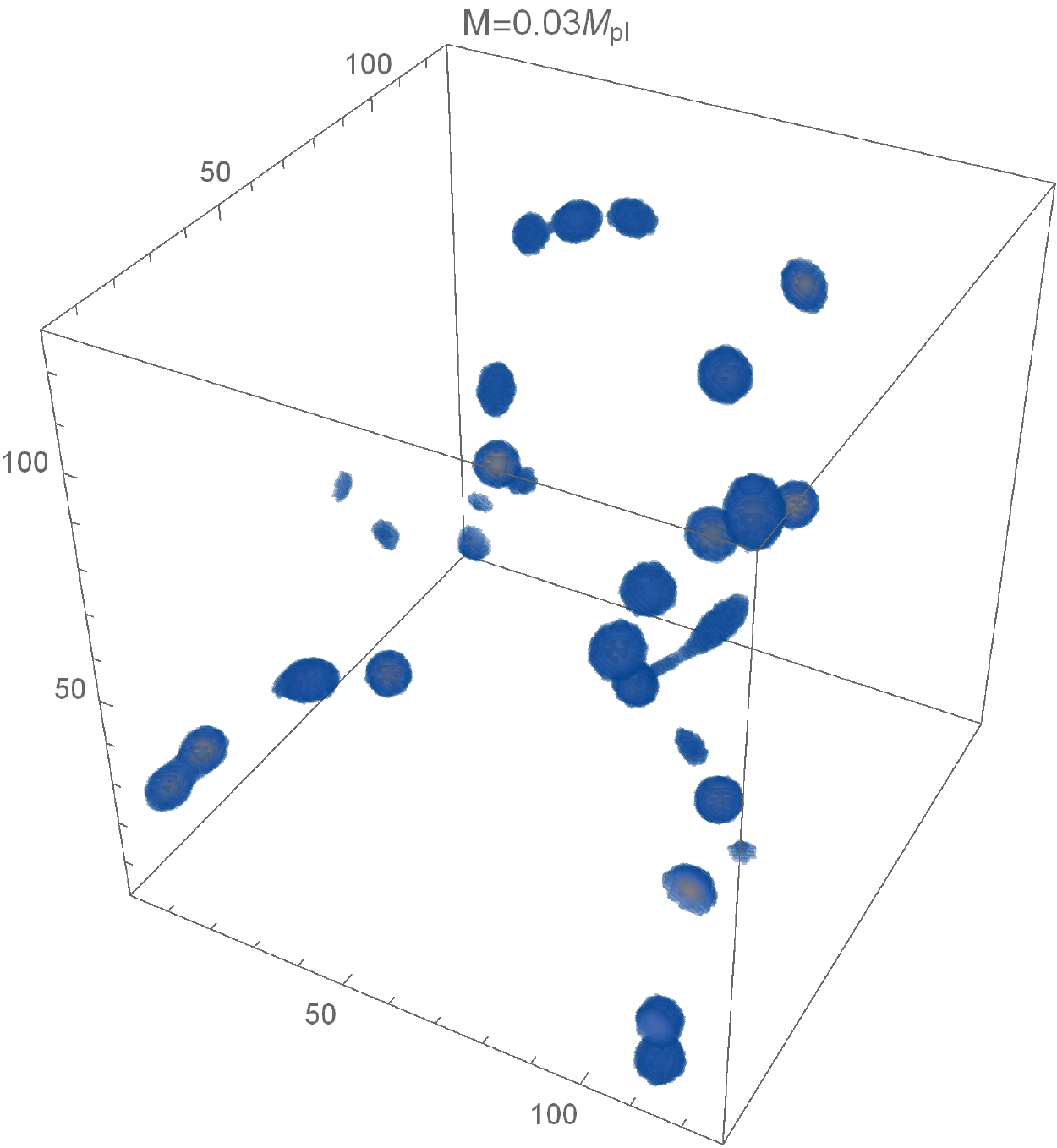}
\caption{
Snapshots of the energy density at the end of our simulations for the cuspy potentials~\eqref{eq:potentials}
with $p=1$ (top-left panel), $p=2/3$ (top-middle panel) and $p=2/5$ (top-right panel),
and for the potentials~\eqref{eq:pot2} with $M=0.01M_{\mathrm{pl}}$ (bottom-left panel),
$M=0.02M_{\mathrm{pl}}$ (bottom-middle panel) and $M=0.03M_{\mathrm{pl}}$ (bottom-right panel)
in the case of $p=1$.
The energy density contours are taken at $\rho=20\langle\rho\rangle$.
}
\label{fig:osc}
\end{figure*}

As linear fluctuations grows exponentially, 
the linear approximation becomes invalid soon.
We have to consider the back-reaction from fluctuations, particle re-scattering and condensate as described in Refs.~\cite{Kofman:1997yn,Greene:1997fu}.


The EoS parameter can help us understand the nonlinear evolution of fluctuations,
which is defined as
\begin{equation}
\label{eq:EoS}
    \omega\equiv\frac{\langle P\rangle }{\langle \rho\rangle}=\frac{\langle\dot{\phi}^2/2-(\nabla\phi)^2/6a^2-V(\phi)\rangle}{\langle \dot{\phi}^2/2+(\nabla\phi)^2/2a^2+V(\phi)\rangle }\,,
\end{equation}
where $\rho$ and $P$ are the energy density and pressure of the field respectively.
The top panels of Fig.~\ref{fig:omega} show the time evolution of the EoS parameters.
It can be described by three successive phases.
In the first phase, the EoS parameter oscillates between $-1$ and $1$.
It is a stage of linear parametric resonance, which we have analyzed in Subsec.~\ref{subsec:la}.
In the second phase from $t_{br}$ to $t_{osc}$, the amplitude of $\omega$ oscillations decreases and approaches zero.
It is a stage of nonlinear re-scattering, in which the zero mode of fluctuations decays rapidly
and higher momentum modes grow rapidly.
Such a phase is more violent in the infinite cuspy models, compared to the finite cuspy model.
Moreover, the time average of $\omega$ over oscillations is negative in the first two phases.
In the third phase, $\omega$ oscillations decay towards zero.
Since there is some energy stored in relativistic modes outside oscillons,
actually $\omega$ is not exactly zero.
It is a stage of oscillon formation, which we shall consider in the next subsection.

The bottom panels of Fig.~\ref{fig:omega} show the evolution of the energy density of the field.
We can see that the time average of the energy density over oscillations
is well described by a power law of $a$ with index of $(-6p)/(2+p)$ in the linear resonance phase.
This is because the time average of the EoS parameter is given by $(p-2)/(p+2)$
when the field oscillates around of the minimum of its potential with a power-law form~\cite{Johnson:2008se,Turner:1983he}.

\subsection{Oscillon formation}
The nonlinear evolution of fluctuations is followed by oscillon formation
and ultimately oscillons dominate the Universe.
The so-called oscillons are localized, compact and long-living objects from the nonlinear solution of a scalar field~\cite{Copeland:1995fq,Gleiser:2008ty,Amin:2010jq,Amin:2013ika}.
Such oscillons can form during preheating after inflation~\cite{Broadhead:2005hn,Farhi:2007wj,Amin:2010dc,Gleiser:2011xj,Amin:2011hj,Lozanov:2016hid,Lozanov:2017hjm,Hasegawa:2017iay}.
Analytic solutions of stable oscillons in the one-dimensional signum-Gordon model is obtained in Refs.~\cite{Arodz:2007jh,Arodz:2011zm},
in which the potential is the same as~\eqref{eq:potentials} with $p=1$.
Generally speaking, oscillons can form if the scalar potential is quadratic near its minimum and flattens away from it,
i.e., the ``opening up" condition~\cite{Amin:2010jq}.
When oscillons are well-separated,
the oscillon field profile is approximately written as
\begin{equation}
\label{eq:osc1}
    \phi(t,\mathbf{x})=\sum_{n=1}^{\infty}\Phi_{n}(\mathbf{x})\cos(n\omega_{\phi} t)\,,
\end{equation}
where $\Phi_n(\mathbf{x})$ are positive localized functions that fall off quickly far from the center of oscillons.
The profile is oscillating at the frequencies of $n\omega_{\phi}$.
The ansatz~\eqref{eq:osc1} with $n=1$ captures the dominant oscillating mode of oscillons.
The even order harmonics in~\eqref{eq:osc1} are absent if the potential is symmetric under $\phi \to -\phi$.
Typically oscillons last for millions of oscillations
and then decay through classical or quantum radiation~\cite{Salmi:2012ta}.
Once oscillon formation is completed, the energy stored in oscillons is approximately constant until they decay.
So the average energy density is proportional to the number density of oscillons like a perfect fluid with $\omega=0$.

It is found in Ref.~\cite{Zhou:2013tsa} that an isolated spherically-symmetric oscillon does not radiate GWs and
the GW emission generated between oscillons is also small.
Therefore, the oscillon-dominated phase itself does not generate significant GWs.
However, oscillon formation can provide a significant GW source.
We now simulate numerically oscillon formation in the models of \eqref{eq:potentials} and \eqref{eq:pot2}
with different values of $M$.

Snapshots of the energy density are plotted in Fig.~\ref{fig:osc}
at the moment when the energy density of GWs does not grow significantly.
The energy density in the colored region is larger than $20$ times the average energy density.
From the top panels of Fig.~\ref{fig:osc}
we can see oscillons with different sizes form in the cuspy potentials.
There are more small-size oscillons produced in the infinite cuspy models than in the finite cuspy model.
As shown in the bottom panels of Fig.~\ref{fig:osc}, the number density of oscillons is small in the case of $M=0.03M_{\mathrm{pl}}$,
compared to the cases of $M=0.01M_{\mathrm{pl}}$ and $M=0.02M_{\mathrm{pl}}$.
The parameter $M$ characterizes the smoothness of the potential around the origin.
Therefore, the cuspy potentials trigger more oscillons than a smooth potential.

\section{Observational Implications}
\label{sec:stoGW}
As expected, a stochastic background of GWs is generated during oscillations of the inflaton after inflation,
which can be detected by the future space-based or ground-based interferometers.
In this section, we shall use the configuration-space method described in Sec.~\ref{sec:nummet}
to numerically simulate the energy spectrum of GWs in the models of \eqref{eq:potentials} and \eqref{eq:pot2}.
We stop the simulation when the energy spectrum of GWs does not grow significantly.

\begin{figure*}[t]
\includegraphics[width=2.7in]{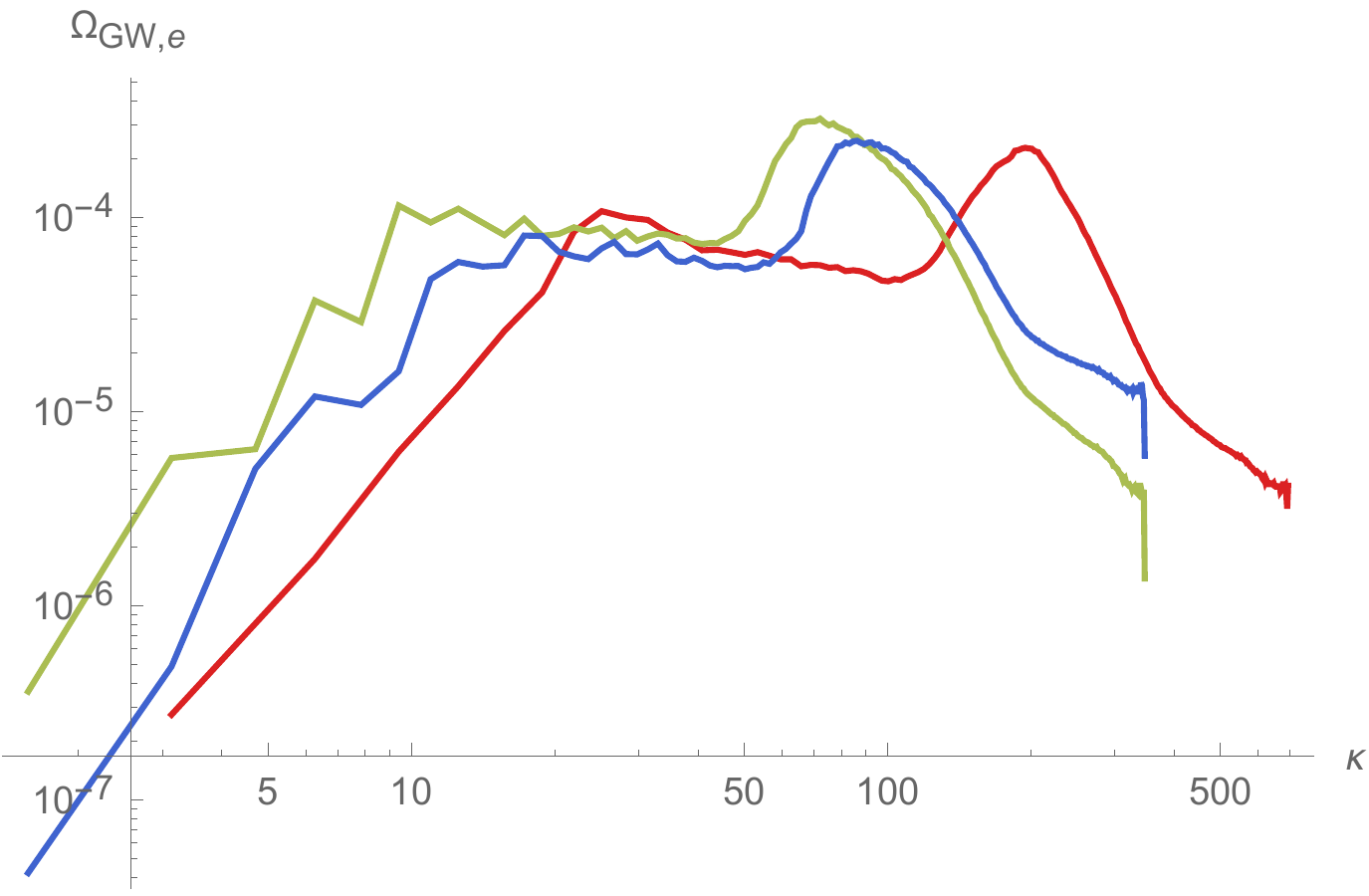}
\raisebox{1.1\height}{\includegraphics[height=0.5in]{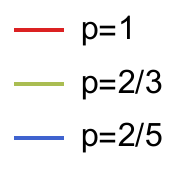}}
\includegraphics[width=2.7in]{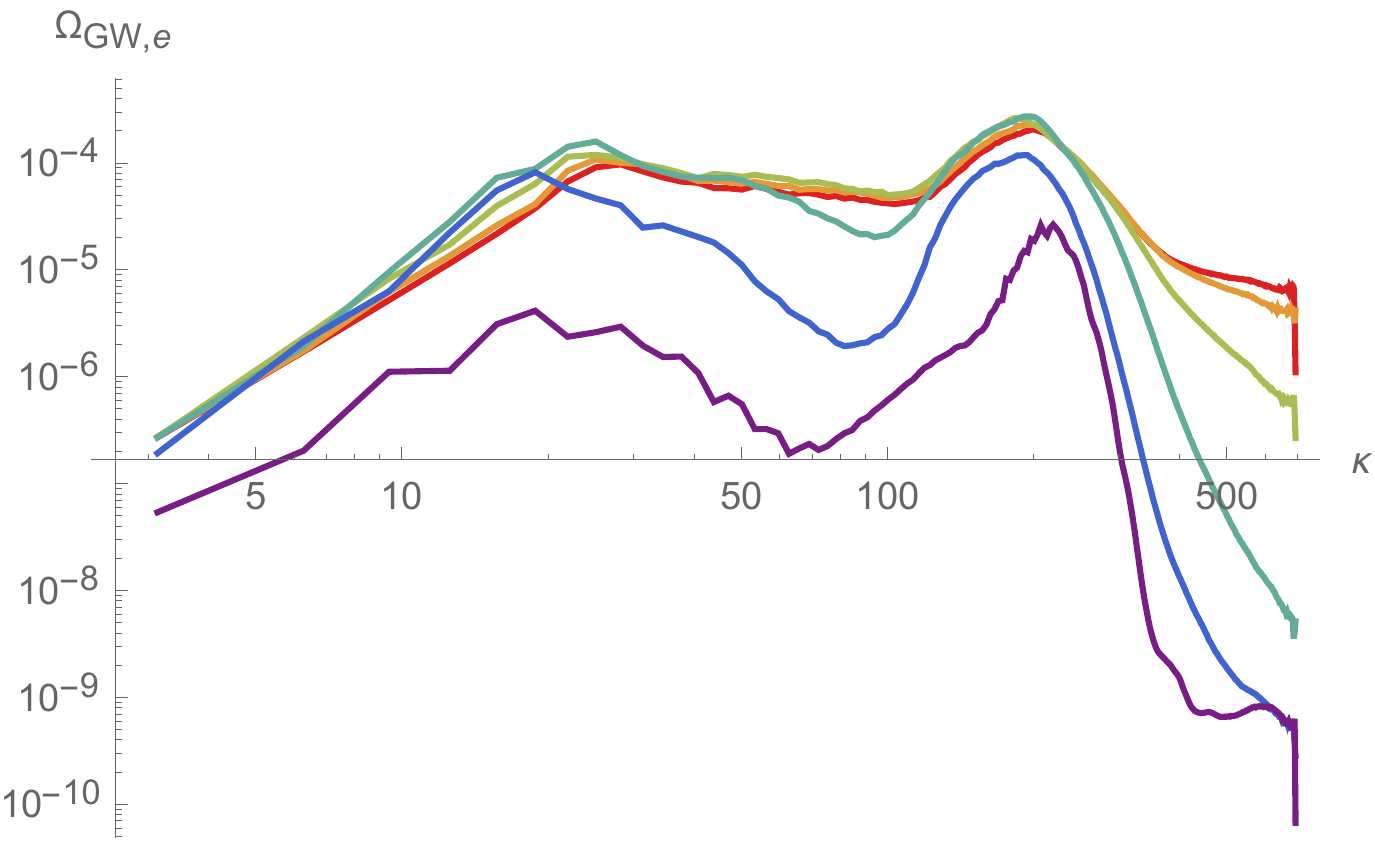}
\raisebox{0.3\height}{\includegraphics[height=1in]{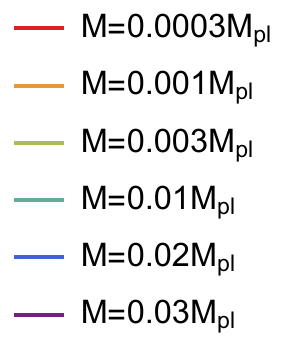}}
\caption{
Energy spectra of GWs at the end of our simulations for the cuspy potentials~\eqref{eq:potentials}
with $p=1$ (red curve), $p=2/3$ (green curve), $p=2/5$ (blue curve) in the left panel,
and for the potential~\eqref{eq:pot2} with $p=1$ and different $M$ in the right panel.
}
\label{fig:GW}
\end{figure*}

The left panel of Fig.~\ref{fig:GW} shows $\Omega_{\mathrm{GW},e}$ at the end of the simulation
in the cuspy potentials with $p=1$ (red curve), $p=2/3$ (green curve) and $p=2/5$ (blue curve),
where the subscript $e$ denotes quantities evaluated at the end of the simulation.
Such cuspy potentials in general predict the energy spectra with the same double-peak structure,
in which the right peak is slightly higher than the left one.
It seems difficult to distinguish these models by detecting the energy spectra of GWs.
Since the smoothness of the potential~\eqref{eq:pot2} around the origin weakens oscillon formation,
the energy spectrum of GWs is suppressed in the large $M$ case, as shown in the right panel of Fig.~\ref{fig:GW}.
When $M\le0.001M_\mathrm{pl}$, the predicted spectrum becomes insensitive to the values of $M$.
Therefore, the cuspy potentials~\eqref{eq:potentials} are well approximated by the potentials~\eqref{eq:pot2} with $M = 0.001M_\mathrm{pl}$.

\begin{figure*}[t]
\includegraphics[width=2.2in]{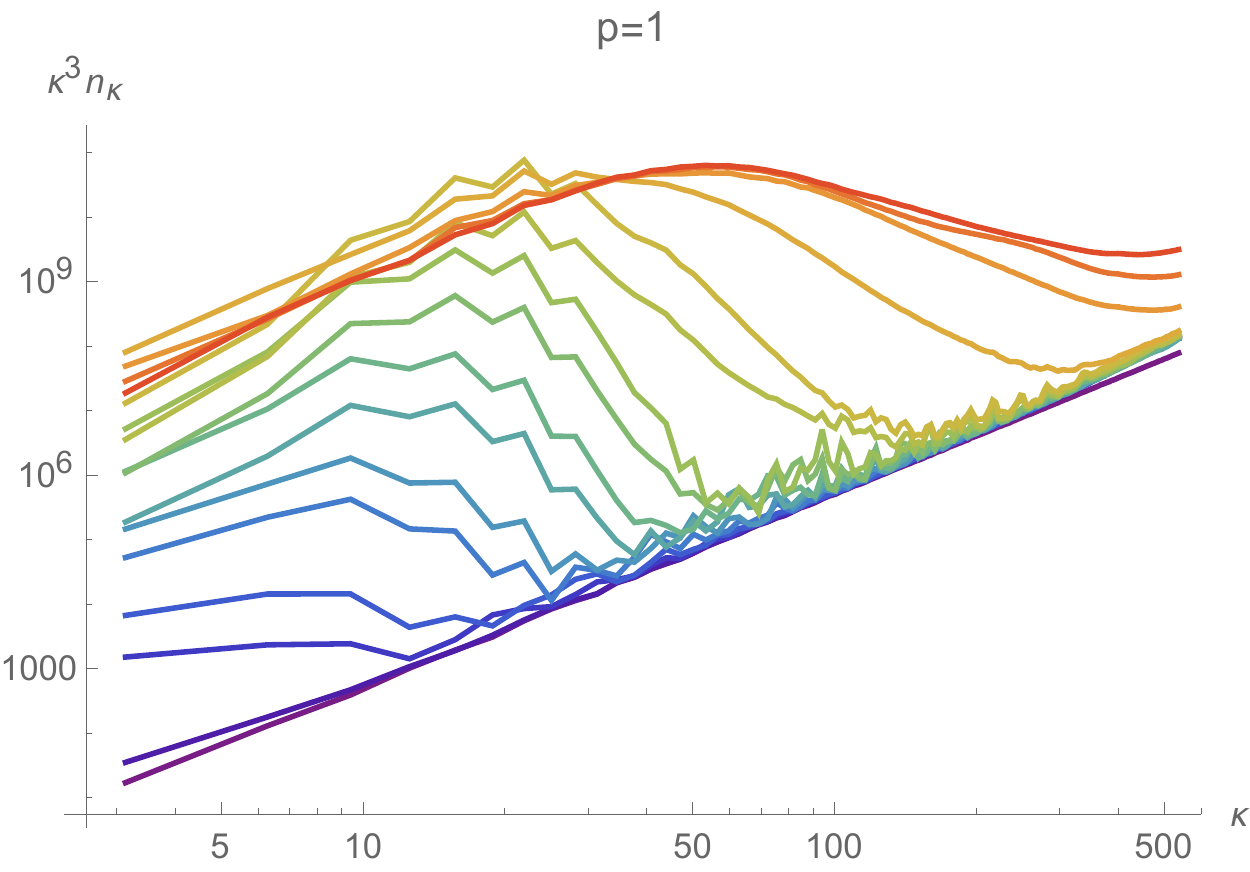}
\includegraphics[width=2.2in]{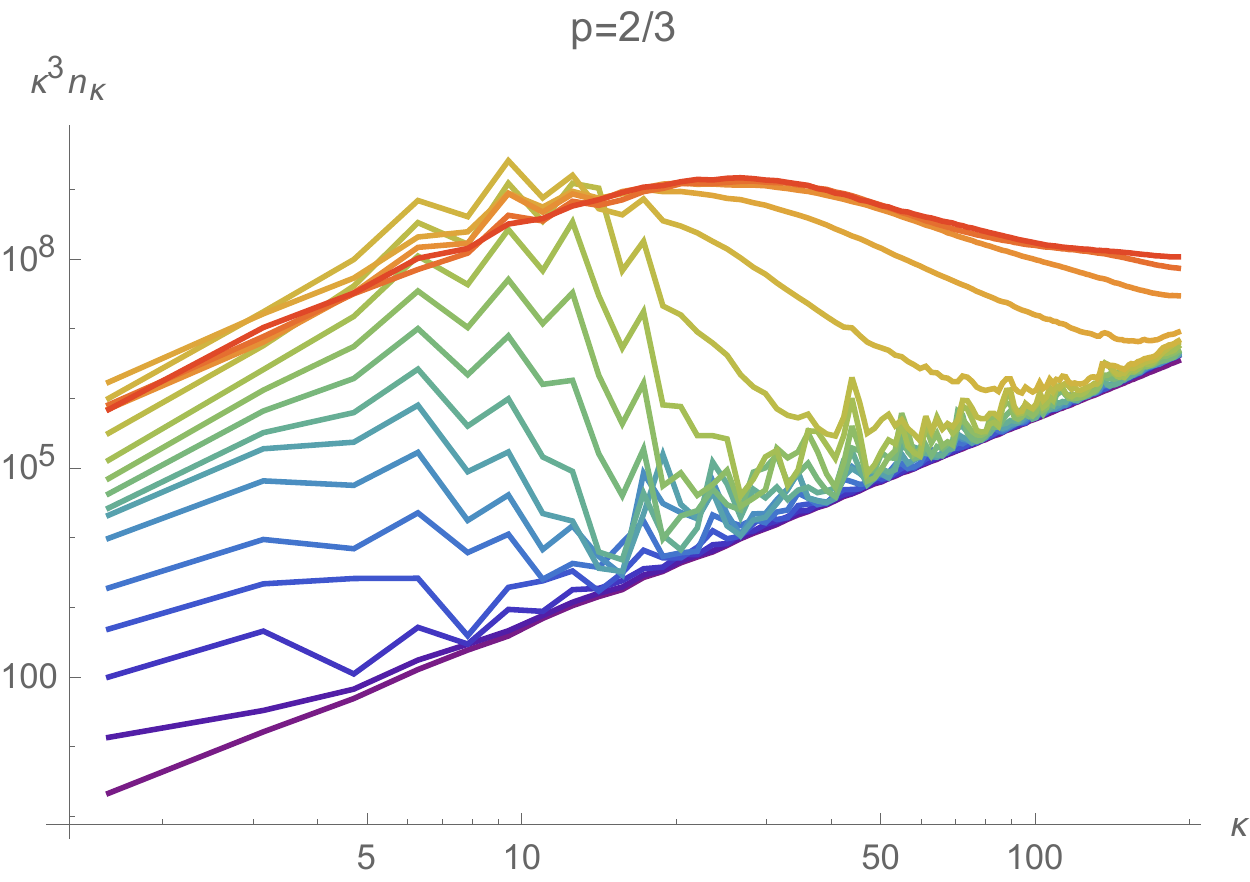}
\includegraphics[width=2.2in]{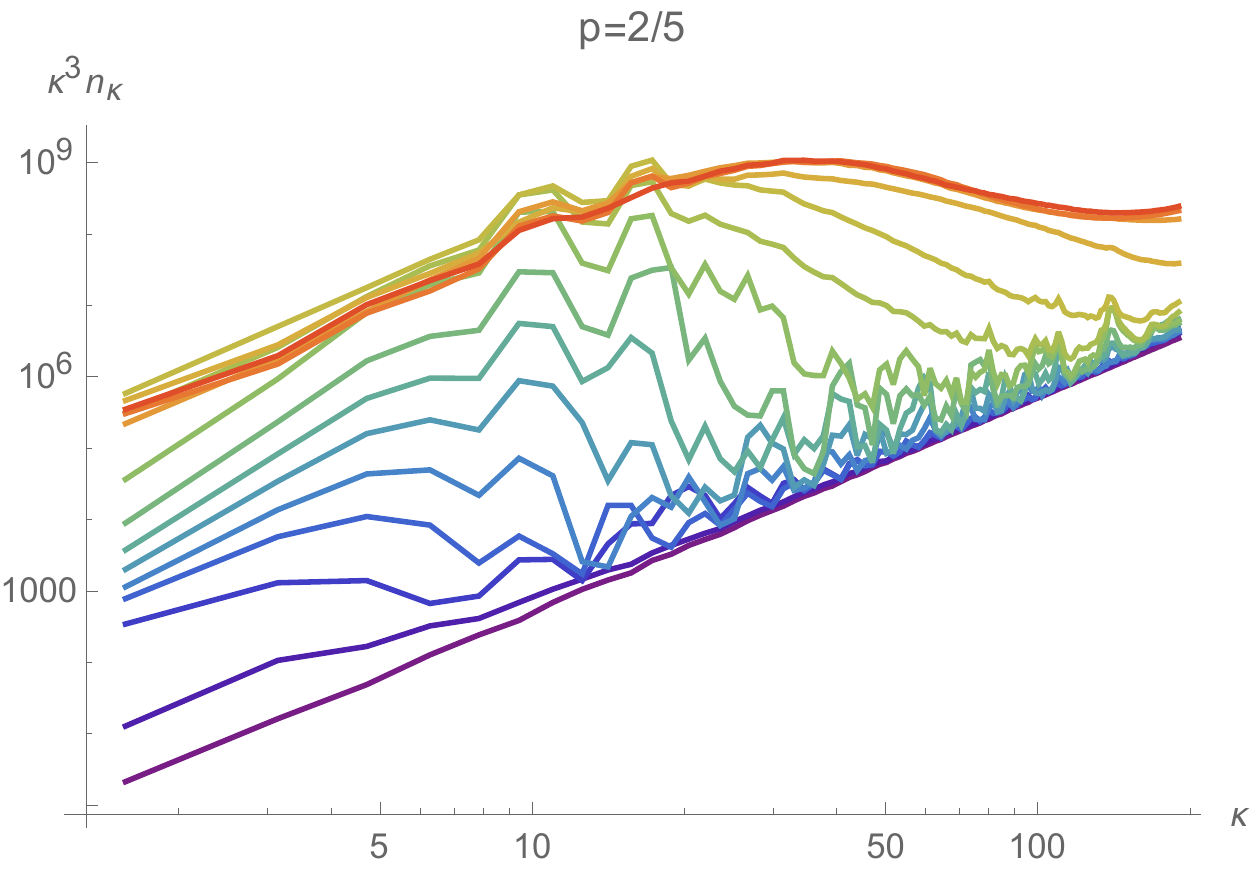}
\caption{
Evolutions of the energy density spectrum of the field for the cuspy potentials~\eqref{eq:potentials}
with $p=1$ (left panel), $p=2/3$ (middle panel) and $p=2/5$ (right panel).
The yellow line corresponds to a turning point $a(t)=7.30$, $3.95$ and $3.56$ for $p=1$, $2/3$ and $2/5$, respectively.
}
\label{fig:nspectrum}
\end{figure*}

It is an interesting feature that in our models the predicted energy spectrum of GWs has double peaks,
which is very distinct from that of other models.
Therefore, our models can be distinguished from the production of GWs produced from preheating by future GW detectors.
Recently, the production of GWs during parametric resonance is studied in standard preheating scenarios
with quadratic and quartic potentials~\cite{Figueroa:2017vfa}.
For some choices of the resonance parameter,
one also finds a GW spectrum with multiple peaks due to nonlinear effects~\cite{Figueroa:2017vfa}.
However, in our models the double peaks are due to copious oscillon formation
and the shapes are independent of the model parameters.

As discussed in~\cite{Liu:2017hua},
the left peak in the energy spectrum of GWs mainly arises from the rapid growth of fluctuations in the linear stage.
Fig.~\ref{fig:nspectrum} shows the evolution of the energy density spectrum of the field in the models~\eqref{eq:potentials}
with $p=1$ (left panel), $p=2/3$ (middle panel) and $p=2/5$ (right panel), where
\begin{equation}
\label{eq:nspe}
    k^{3}\omega_k n_{k}=\frac{1}{2}k^{3}\left[|\partial_{\tau}(a\delta\phi_{k})|^{2}+\omega^2_{k}|a\delta\phi_{k}|^{2}\right]\,.
\end{equation}
The evolution of the spectrum goes through three different phases, i.e.,
linear growth, nonlinear re-scattering and oscillon formation.
In the first phase, the small-$k$ modes in the resonance bands exponentially grow due to the cusp of the potential
until the turning point $a(t)=7.30$, $a(t)=3.95$ and $a(t)=3.56$
for $p=1$, $p=2/3$ and $p=2/5$, respectively,
so that the spectrum achieves a peak at small $k$ (see Fig.~\ref{fig:nspectrum}).
This leads to the left peak in the energy spectrum of GWs,
which is characteristic of the cuspy potentials.
Due to the subsequent nonlinear re-scattering the small-$k$ modes begin to drop
and the large-$k$ modes continue to grow until oscillons are properly formed.
It implies that the energy flows from the small-$k$ modes to the large-$k$ modes,
as discussed in detail in the standard preheating scenario~\cite{GarciaBellido:2007af}.
The growth of the large-$k$ modes in the oscillon formation stage leads to the right peak in the energy spectrum of GWs.
As expected, the forming of oscillons leaves imprints in the energy spectrum of GWs.
Our simulations indicate that the frequency of the right peak in the energy spectrum of GWs
is twice the dominant oscillating frequency of oscillons.
This is argued as follows.
The dominant oscillating mode of the oscillon~\eqref{eq:osc1} reads
\begin{equation}
\label{eq:EoMosc}
\phi(t,\mathbf{x})=\Phi_1(\mathbf{x})\cos(\omega_{\phi} t)\,.
\end{equation}
Since the source term in~\eqref{eq:EoMuij} contains only quadratic terms of $\partial_{i}\phi$,
it implies that the source term contains two frequencies with $2\omega_{\phi}$ and $0$,
\be
\label{eq:EoMsource}
T_{ij} \sim \cos(2\omega_{\phi} t) + 1 \,.
\ee
The characteristic frequency with $2\omega_{\phi}$ corresponds to the right peak frequency of $\Omega_{\mathrm{GW}}$.
The energy spectrum of GWs with zero frequency is highly suppressed from the relation $\Omega_{\mathrm{GW}} \propto k^3$.
In principle, the source term contains higher order harmonics of the oscillon.
Compared to the leading order mode, the higher order mode contributions to $\Omega_{\mathrm{GW}}$ can be neglected.

\begin{figure*}[t]
\includegraphics[width=6in]{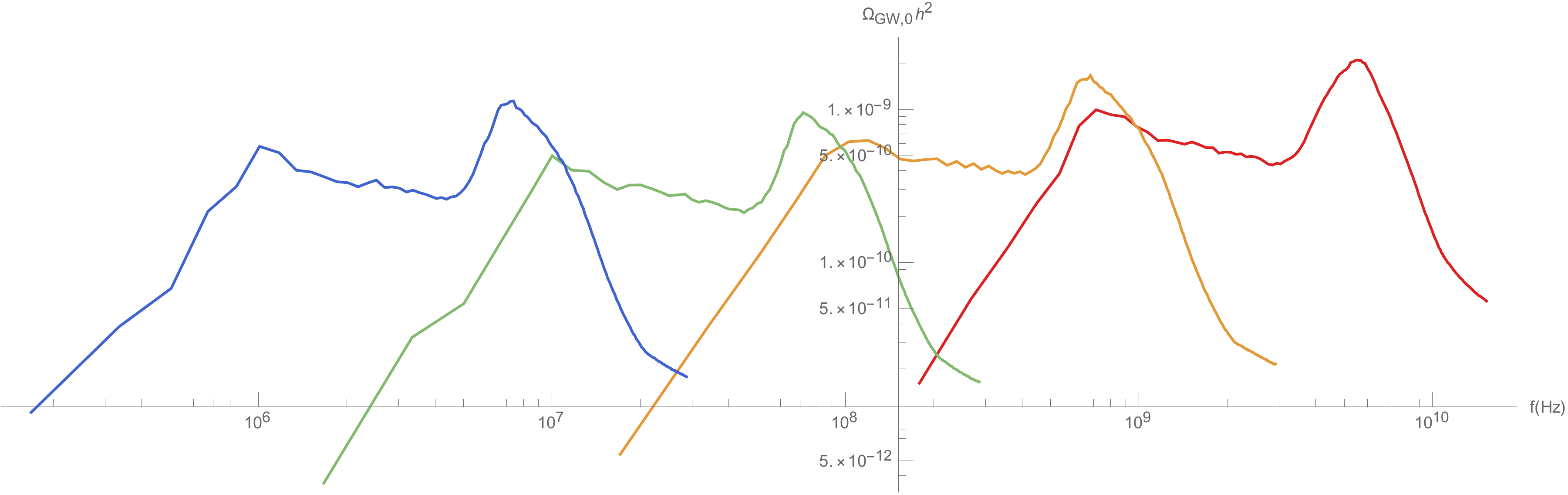}
\raisebox{0.8\height}{\includegraphics[height=0.8in]{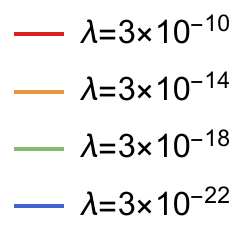}}
\caption{
Energy spectra of GWs today for the linear potential with $\lambda=3\times10^{-20}$ (left)
through to $3\times 10^{-10}$ (right).
}
\label{fig:scaleGW}
\end{figure*}

In order to compare the predicted energy spectrum with the sensitivity curves of the future detectors,
let us estimate the present value of $\Omega_{\mathrm{GW}}$ and the corresponding frequency $f$.
We define $t_e$ as the time at the end of the simulation,
$t_*$ the time when thermal equilibrium is established,
$t_0$ the present time,
and $g$ the effective number of ultrarelativistic degree of freedom.
Since the energy density of radiation evolves as $\rho_{r}\propto g^{-1/3}a^{-4}$,
the energy spectrum at the present time is related to that at the end of the simulation as
\begin{equation}
\label{eq:GWnow}
\Omega_{\mathrm{GW},0}
=\Omega_{r,0}\left(\frac{g_{0}}{g_{*}}\right)^{1/3}\left(\frac{a_{e}}{a_{*}}\right)^{1-3\omega}\Omega_{\mathrm{GW},e}\,,
\end{equation}
where $\Omega_{r,0}$ is the density fraction of radiation today,
and $\omega$ is the effective EoS parameter between $t_{e}$ and $t_{*}$.
The present value of frequency $f$ is
\begin{equation}
\label{eq:fnow}
f
\simeq  \frac{k}{a_{e}\rho_{e}^{1/4}} \left(\frac{g_0}{g_*}\right)^{1/12} \left(\frac{a_{e}}{a_{*}}\right)^{(1-3\omega)/4}4 \times 10^{10}\;\mathrm{Hz}\,.
\end{equation}
We assume that reheating is completed at the end of the simulation.
It means that $a_e=a_*$.
In our calculations, we use $g_*/g_0=100$.

As discussed in~\cite{Easther:2006gt},
if inflation happens at lower energy scales, the GW energy density will be diluted less by the expansion till the present time.
On the other hand, lowering the energy scale of inflation leads to less efficient sources of gravitational radiation during preheating.
These two effects roughly cancel each other and hence the energy spectrum of GWs does not depend strongly on the energy scale of inflation.
Fig.~\ref{fig:scaleGW} shows the energy spectrum of GWs for different energy scales of inflation with the cuspy potentials~\eqref{eq:potentials}.
As expected, the peak frequency scales with the energy scale of inflation
while the amplitude is independent of this scale.
Actually, the comoving wavenumber $k$ in~\eqref{eq:fnow} can be estimated by $\sqrt{\lambda/\phi_i}$
as discussed in Sec.~\ref{sec:dyn}.
Therefore, reducing $\lambda$ pushes the signature towards lower frequencies.

However, in the smooth potential~\eqref{eq:pot2} with large values of $M$,
the energy scale of inflation affects the amplitude of the energy spectrum of GWs.
Field fluctuations are initialized by quantum vacuum fluctuations in our simulations,
as described in Sec.~\ref{sec:nummet}.
The initial values depend on the energy scale of inflation.
Lowering the energy scale leads to smaller initial values of fluctuations.
As linear fluctuations grow rapidly, the amplitude of field oscillations decrease.
When $\phi < M$, the growth of fluctuations is suppressed by
a quadratic potential.
Therefore, the energy spectrum of GWs is suppressed by a lower energy scale of inflation
in the smooth potential~\eqref{eq:pot2} with large values of $M$.
For example, choosing $M=0.02M_\mathrm{pl}$ we find that
the energy spectra of GWs peak at around $\Omega_{\mathrm{GW},e} \sim 10^{-4}$ in the case of $m=1.22 \times 10^{-4} M_\mathrm{pl}$
while $\Omega_{\mathrm{GW},e} \sim 10^{-20}$ in the case of $m=1.22 \times 10^{-10}M_\mathrm{pl}$.
This differs from the cuspy models in which the amplitude of the energy spectrum of GWs is independent of the energy scale of inflation.

\begin{figure}[h]
\includegraphics[width=3.2in]{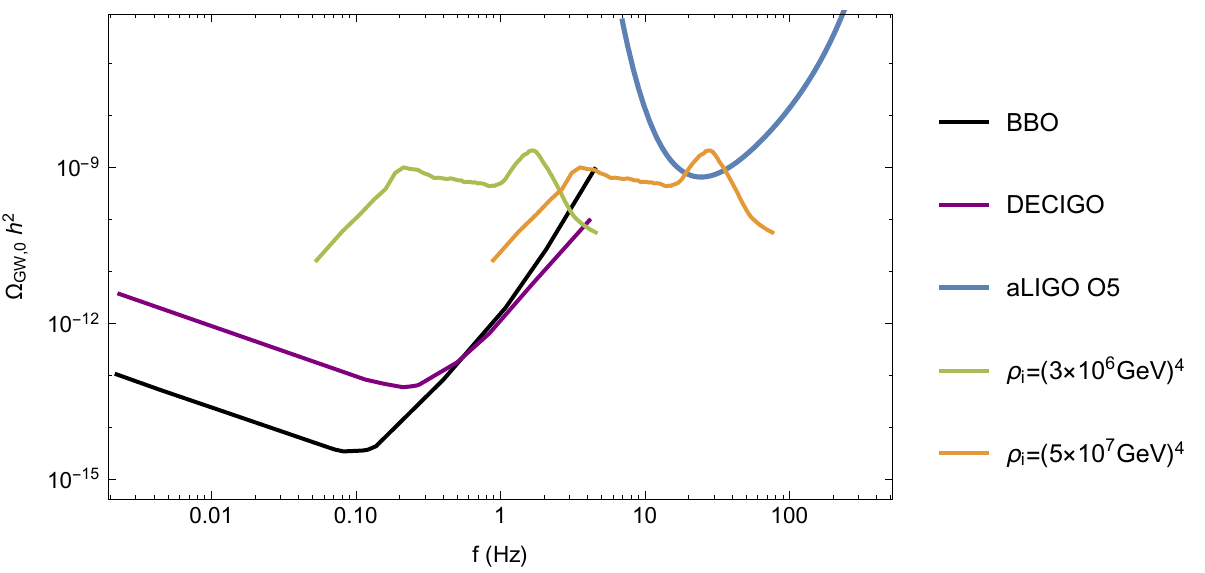}
\caption{
Energy spectra of GWs today, predicted by the linear potential
with the energy scales of $3\times10^{6}\;\mathrm{GeV}$ (green)
and $3\times10^{7}\;\mathrm{GeV}$ (orange).
The black, red, and blue curves are the expected sensitivity curves of BBO, DECIGO and aLIGO-Virgo detectors, respectively.
}
\label{fig:observe}
\end{figure}

Since GWs are generated causally within the Hubble volume at that time
and simple inflationary models typically happens at the GUT scale,
the typical wavelength of these GWs is considerably shorter than LIGO scales.
For example, in the single-field slow-roll inflationary model~\eqref{eq:potentials} with $p=1$,
if $\lambda\approx 3\times 10^{-10}$ is fixed by the amplitude of the primordial curvature perturbations
$A_s=2.2\times 10^{-9}$,
the peak frequency of GWs today is fixed to be $f\sim 10^9$ Hz,
many orders of magnitude beyond the frequencies that can be reached by ground-based GW detection experiments.
If the model parameter $\lambda$ is not fixed by the amplitude of the primordial curvature perturbations,
advance LIGO (aLIGO) whose sensitivity is expected to be significantly improved,
allows us to possibly observe GWs produced during oscillations of the inflaton after inflation.
For example, in the hybrid inflationary scenario~\cite{Linde:1993cn},
$\lambda$ becomes essentially a free parameter because $\phi$ is not necessarily the inflaton itself.
In this case in Fig.~\ref{fig:observe} we have plotted the present-day energy spectra of GWs
produced during oscillon formation in the linear potential model~\eqref{eq:potentials}
with $\lambda = 1.88 \times 10^{-43}$ (orange) and $\lambda = 2.43 \times 10^{-48}$ (green).
We can see that the peaks lie above the expected sensitivity curve of the fifth observing run (O5) of
the aLIGO-Virgo detector network~\cite{TheLIGOScientific:2016wyq}.
As shown in Fig.~\ref{fig:observe}, there are two peaks in the energy spectrum of GWs,
which differ from other spectra of GWs produced during preheating.
A detection of the second peak may require corroboration from low-frequency GW detectors such as the Big Bang Observatory.


In the hybrid inflationary scenario, the energy scale of inflation ranges from the GUT scale all the way down to the electroweak scale.
Consequently, oscillon formation generates a stochastic background of GWs with
a typical frequency today of the order of $10^{-3} - 10^{9}$ Hz.
Present ground-based~\cite{TheLIGOScientific:2016wyq} detectors work at frequencies of $1-10^3$ Hz
and planned space-based GW detectors~\cite{Bartolo:2016ami,Guo:2018npi} work
at frequencies of $10^{-3}-1$ Hz,
which provide a possibility to detect low-frequency GW signals.
It is proposed in~\cite{Li:2017jcz} that a coupling system between Gaussian type-microwave photon flux,
static magnetic field and fractal membranes can be used to detect high-frequency GWs in the microwave band.
This opens a new window of high-frequency GW detection.

\section{Conclusions and Discussions}
\label{sec:condis}
We have investigated the effects of the cuspiness of the potentials on the production
of GWs in the oscillon preheating scenario.
For comparison we turn to the more general form of the potentials~\eqref{eq:pot2},
which can well approximate the cuspy potentials when $\phi/M$ is large.
Our simulations indicate that the predicted energy spectrum of GWs
becomes insensitive to the values of $M$ when $M \le 0.001M_\mathrm{pl}$.
We find the potential with a cusp at its minimum yields stronger GW signals than smooth potentials.
Due to the cusp, the oscillating behavior of the inflaton is different from that of smooth potentials.
The nonsmooth oscillations can trigger a significant amplification of the field fluctuations,
so that oscillons copiously form, which leads to a significant GW signal.
Moreover, in the cuspy potentials the amplitude of the energy spectrum of GWs is independent of the energy scale of inflation,
while in the smooth potentials~\eqref{eq:pot2} with large values of $M$,
lowering the energy scale of inflation reduces power in the energy spectrum of GWs today.
By varying the parameter $M$, we find that cuspy potentials yield stronger signals
of gravitational waves and the generation of gravitational waves disappears for smooth potentials.

We have studied the dynamics of the oscillon preheating, which is described by three successive phases,
i.e., linear parametric resonance, nonlinear re-scattering and oscillon formation.
In the first stage, the small-$k$ modes in the resonance bands exponentially grow due to the cusp of the potential until the turning point.
This leads to the left peak in the energy spectrum of GWs.
The effective EoS parameter oscillates from $-1$ to $1$.
In the second stage, the energy flows from the small-$k$ modes to the large-$k$ modes.
The growth of the large-$k$ modes in the subsequent stage leads to the right peak in the energy spectrum of GWs.
The effective EoS parameter tends to zero.
It implies the Universe goes into a quasi matter-dominated stage right before the transition to the radiation-dominated stage.


In our analysis we have neglected the interactions between the inflaton $\phi$ and other matter fields.
If a matter field $\chi$ is coupled to the inflaton,
broad parameter resonance actually leads to a fast growth of the $\chi$ fluctuations.
As found in Ref.~\cite{Moghaddam:2015ava},
an efficient parameter resonance can occur during preheating for a cuspy potential
with a coupling term $\frac{1}{2}g^{2}\phi^{2}\chi^{2}$.
However, our numerical simulations confirm that the growth of the inflaton
fluctuations themselves triggered by the cusp in its potential is more
effective than that of the field $\chi$ by parametric resonance.
For example, if the coupling constant $g$ is chosen as $g^{2}/\lambda = 3\times10^{4}, 10^{6}, 10^{8}, 10^{10}$
in the cuspy potentials~\eqref{eq:potentials},
we find the field $\chi$ has little impact on the evolution of the inflaton fluctuations and oscillon formation.
Therefore, GWs are sourced mainly by the inflaton fluctuations,
even if a parametric resonance for the field $\chi$ occurs in our models.

Oscillon formation is completed at the end of the simulation.
After that the oscillons survive until the Universe is heated due to their decay.
Since density perturbations grow in the oscillon-dominated phase,
the collapse of oscillons can lead to copious production of primordial black holes~\cite{Cotner:2018vug}.
The dynamics of the latter can provide yet another source of GWs.

\begin{acknowledgements}
This work is supported in part by the National Natural Science Foundation of China Grants
No.11435006, No.11575272, No.11690021, No.11690022, No.11851302 and No.11821505,
in part by the Strategic Priority Research Program of the Chinese Academy of Sciences Grant No.XDB23030100,
No.XDA15020701 and by Key Research Program of Frontier Sciences, CAS.
GS is supported in part by the DOE grant DE-SC0017647 and the Kellett Award of the University of Wisconsin.
\end{acknowledgements}

\bibliographystyle{apsrev}
\bibliography{cusp}

\end{document}